\documentclass[a4paper,11pt]{article}
\usepackage{bm,mathrsfs,amsmath,amsthm,amssymb,amscd,longtable,array,graphicx}
\newcommand{\nc}{\newcommand}
\nc{\rnc}{\renewcommand}
\nc{\nn}{\nonumber}
\nc{\der}{{\partial}}
\rnc{\Im}{{\rm{Im}\,}}
\rnc{\Re}{{\rm{Re}\,}}
\nc{\db}{\displaybreak[0]\\}
\nc{\bra}{\langle}
\nc{\ket}{\rangle}
\nc{\bs}{\boldsymbol}

\newtheorem{theorem}{Theorem}[section]

\newtheorem{proposition}[theorem]{Proposition}

\theoremstyle{definition}
\newtheorem{definition}[theorem]{Definition}

\numberwithin{equation}{section}

\numberwithin{equation}{section}

\begin{document}%
%
\title{Symmetric functions and
wavefunctions of the six-vertex model
by Izergin-Korepin analysis}

\author{
Kohei Motegi \thanks{E-mail: kmoteg0@kaiyodai.ac.jp}
\\\\
{\it Faculty of Marine Technology, Tokyo University of Marine Science and Technology,}\\
 {\it Etchujima 2-1-6, Koto-Ku, Tokyo, 135-8533, Japan} \\
\\\\
\\
}

\date{\today}

\maketitle

\begin{abstract}
We analyze wavefunctions of the six-vertex model
by extending the Izergin-Korepin analysis on the domain wall
boundary partition functions.
We particularly focus on the case with triangular boundary.
By using the $U_q(sl_2)$ $R$-matrix and a special class of the
triangular $K$-matrix,
we first introduce an analogue of the wavefunctions
of the integrable six-vertex model with triangular boundary.
We first give a characterization of the wavefunctions by
extending our recent work of the Izergin-Korepin analysis
of the domain wall boundary partition function with triangular boundary,
and then determine the explicit form of the symmetric functions
representing the wavefunctions
by showing that it satisfies all the required properties.
We also illustrate the Izergin-Korepin analysis for
the case of ordinary wavefunctions as it is the basic case.
\end{abstract}

\section{Introduction}
The most fundamental and important objects
in statistical physics are partition functions.
Exact computations of partition functions have great impacts
in mathematical physics and mathematics.
The field of integrable lattice models
\cite{Bethe,FST,Baxter,KBI} offers us chances to
compute partition functions exactly,
and have stimulated the advances of
representation theory and algebraic combinatorics.
One of the most basic partition functions called
the domain wall boundary partition function \cite{Ko,Iz}
has become famous in the field of algebraic combinatorics in 1990s,
since its determinant form was realized to be a certain generalization
of the generating function of the enumeration of the alternating sign matrix \cite{Br,Ku1,Ku2,Okada}.
Relation with classical integrable models \cite{FWZ},
orthogonal polynomials \cite{CP},
asymptotics in the thermodynamic limit \cite{KZ,BL,RK} were also investigated.

Various variations of the domain wall boundary partition functions
were introduced. One of the motivations of studying
these variations comes from the expectation that
investigating variations leads us to applications
to the variations of the enumeration of the alternating sign matrices
\cite{Ku2,Okada,Tsuchiya,HK1,HK2}.

In recent years, more generic partition functions which we shall call as
the wavefunctions are attracting attention.
The domain wall boundary partition function can be regarded
as the simplest case of the wavefunctions.
By studying the wavefunctions
of various integrable models and various boundary conditions,
it has been recognized that
their explicit forms are nothing but symmetric functions
such as the Schur, Hall-Littlewood, 
Grothendieck polynoimals and their deformations.
As one of the applications of
these integrable model realizations of symmetric polynomials,
we can discover various new algebraic combinatorial
identities which give substantial generalizations of the
Cauchy, dual Cauchy, Littlewood, Tokuyama formulae and
the Littlewood-Richardson coefficients.
There are now a lot of papers on this subject.
See
\cite{Bogo,BW,BWZ,WZnew,vDE,MS,Motegi,MS2,Korff,GK2,Borodin,BP1,Takeyama,WZ}
for examples which investigate symmetric functions
by using the XXZ model and the $q$-boson model,
and
\cite{BBF,Iv,BBCG,Tabony,BMN,LMP}
by using the free-fermion model in an external field
.

In this paper, we introduce and study an analogue of the wavefunctions.
We investigate the wavefunctions with triangular boundary.
The main motivation of introducing this object is to generalize
combinatorial objects such as the non-intersecting lattice paths and
excited Young diagrams in Schubert calculus \cite{St,Ok,IN1,IN2,HK}.
By translating those combinatorial objects into the language
of integrable lattice models, one finds that
those combinatorial objects use the $q=0$ degeneration of the
$U_q(sl_2)$ $R$-matrix as bulk weights.
From the point of view of quantum integrability,
it is natural to have the following feeling:
why don't we use the $U_q(sl_2)$ $R$-matrix itself as bulk weights?
In our very recent work \cite{Motr},
we analyzed the domain wall boundary partition function
with triangular boundary which is the very first step,
and determined its explicit expression as a certain
symmetric function. The Izergin-Korepin analysis \cite{Ko,Iz,Wheeler}
was crucial to get the explicit form.
In this paper, we extend the Izergin-Korepin analysis
to the wavefunctions, and determine the explicit form
which is an extension of the symmetric function
representing the domain wall boundary partition function.

This paper is organized as follows.
In the next section, we first list the $U_q(sl_2)$ $R$-matrix
and the triangular $K$-matrix which we use in this paper.
Using these $R$-matrix and $K$-matrix,
we introduce the wavefunctions
with triangular boundary in section 3.
In section 4, we make the Izergin-Korepin analysis
and list the properties needed to determine the explicit form
of the wavefunctions.
In section 5, we present the symmetric function which gives the explicit
form of the wavefunctions by showing that it satisfies
all the required properties.
Section 6 is devoted to conclusion.
We illustrate the Izergin-Korepin analysis to
the ordinary wavefunctions in the Apeendix.

\section{The six-vertex model}
In this section, we introduce the $U_q(sl_2)$ $R$-matrix
and the $K$-matrix which will be used as local bulk and boundary pieces
of the wavefunctions with triangular boundary
which will be introduced in the next section.

The most fundamental objects in integrable lattice models
are the $R$-matrix.
In this paper, we use the following $U_q(sl_2)$ $R$-matrix \cite{Dr,J}
\begin{eqnarray}
R_{ab}(u,w)=\left( 
\begin{array}{cccc}
u-tw & 0 & 0 & 0 \\
0 & t(u-w) & (1-t)u & 0 \\
0 & (1-t)w & u-w & 0 \\
0 & 0 & 0 & u-tw
\end{array}
\right), \label{rmatrix}
\end{eqnarray}
acting on the tensor product $W_a \otimes W_b$
of the complex two-dimensional space $W_a$.
Let us denote the orthonormal basis of $W_a$ and its dual as
$\{|0 \rangle_a, |1 \rangle_a \}$ and $\{{}_a \langle 0|, {}_a \langle 1|\}$,
and the matrix elements of the $R$-matrix as
$
{}_a \langle \gamma | {}_b \langle \delta | R_{a b}(u,w)
|\alpha \rangle_a | \beta \rangle_b=[R_{a b}(u,w)]_{\alpha \beta}^{\gamma \delta}
$. The matrix elements of the $R$-matrix are explicitly given as
\begin{align}
{}_a \langle 0| {}_b \langle 0 | R_{a b}(u,w)
|0 \rangle_a | 0 \rangle_b&=u-tw, \\
{}_a \langle 0| {}_b \langle 1 | R_{a b}(u,w)
|0 \rangle_a | 1 \rangle_b&=t(u-w), \\
{}_a \langle 0| {}_b \langle 1 | R_{a b}(u,w)
|1 \rangle_a | 0 \rangle_b&=(1-t)u, \\
{}_a \langle 1| {}_b \langle 0 | R_{a b}(u,w)
|0 \rangle_a |1 \rangle_b&=(1-t)w, \\
{}_a \langle 1| {}_b \langle 0 | R_{a b}(u,w)
|1 \rangle_a | 0 \rangle_b&=u-w, \\
{}_a \langle 1| {}_b \langle 1 | R_{a b}(u,w)
|1 \rangle_a | 1 \rangle_b&=u-tw.
\end{align}

The $R$-matrix \eqref{rmatrix} satsifies the Yang-Baxter relation
\begin{align}
R_{ab}(u,v)R_{ac}(u,w)R_{bc}(v,w)
=R_{bc}(v,w)R_{ac}(u,w)R_{ab}(u,v), \label{yangbaxter}
\end{align}
acting on $W_a \otimes W_b \otimes W_c$.

\begin{figure}[ht]
\includegraphics[width=12cm]{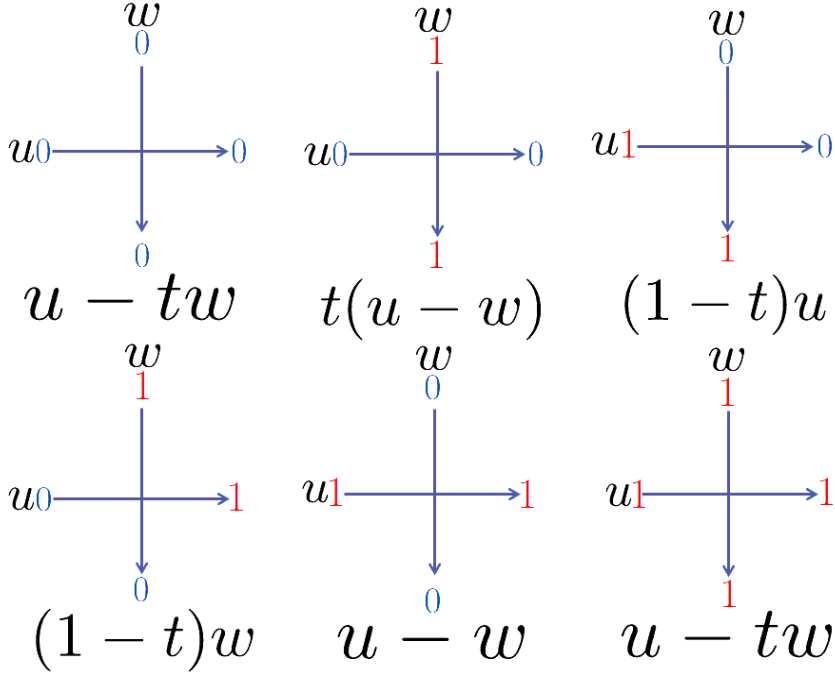}
\caption{The $R$-matrix $R(u,w)$ \eqref{rmatrix}.
We regard that each line is a representation space
and carries a spectral parameter.
In this picture, the horizontal line carries
a spectral parameter $u$, while the vertical line
carries $w$.
}
\label{pictureloperator}
\end{figure}

\begin{figure}[ht]
\includegraphics[width=12cm]{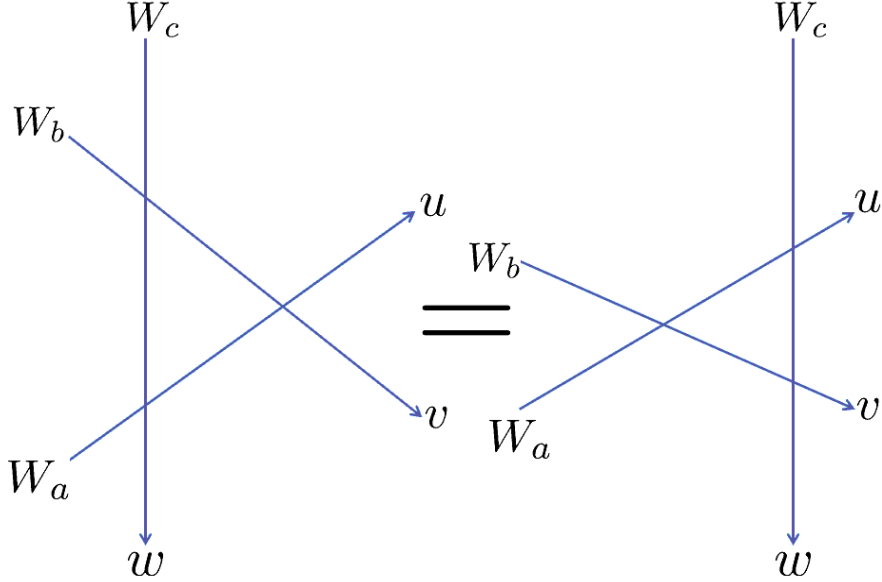}
\caption{The Yang-Baxter relation \eqref{yangbaxter}.
The left and right figure represents
$R_{ab}(u,v)R_{ac}(u,w)R_{bc}(v,w)$
and $R_{bc}(v,w)R_{ac}(u,w)R_{ab}(u,v)$ respectively.
}
\label{pictureyangbaxter}
\end{figure}

See Figures \ref{pictureloperator} and \ref{pictureyangbaxter}
for the graphical description of the $R$-matrix
and the Yang-Baxter relation used in this paper.
The $R$-matrices have origins in statistical physics,
and $| 0 \rangle$ or its dual $\langle 0|$
can be regarded as a spin up state,
while $| 1 \rangle$ or its dual $\langle 1|$
can be interpretted as a spin down state
from the point of view of statistical physics.
We sometimes use the terms spin up states (up spins)
and spin down states (down spins)
to describe states constructed from
$| 0 \rangle$, $\langle 0|$, $| 1 \rangle$ and $\langle 1|$
since they are convenient for the description
of the states.

\begin{figure}[ht]
\includegraphics[width=12cm]{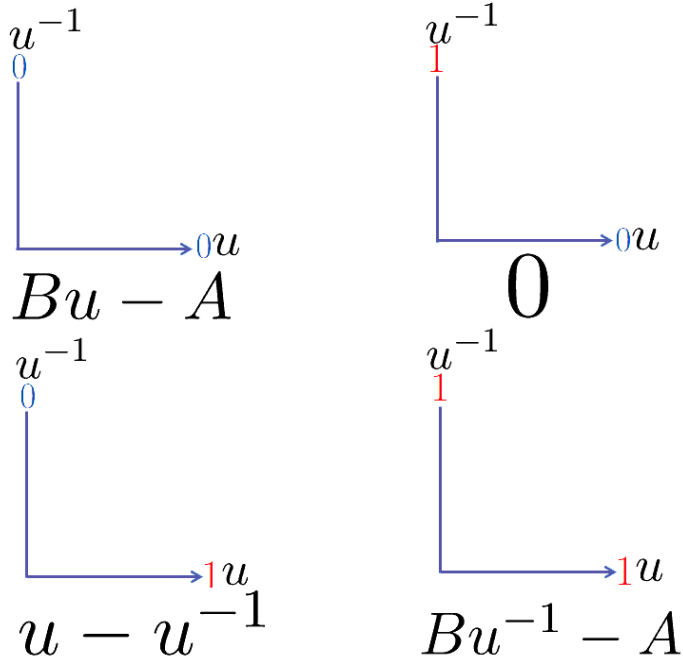}
\caption{The $K$-matrix $K(u)$ \eqref{kmatrix}.
We regard that the horizontal line carries a spectral parameter $u$,
while the vertical line carries its inverse $u^{-1}$.
}
\label{picturekmatrix}
\end{figure}

\begin{figure}[ht]
\includegraphics[width=12cm]{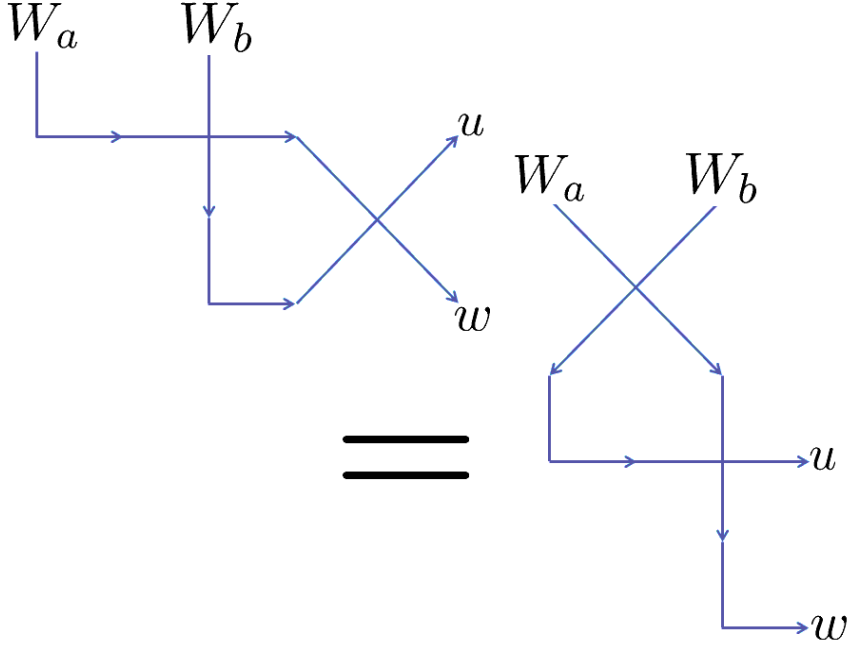}
\caption{The reflection equation \eqref{reflection equation}.
The left and right figure represents
$R_{ba}(u/w)K_b(u)R_{ab}(uw)K_a(w)$
and $K_a(w)R_{ba}(uw)K_b(u)R_{ab}(u/w)$ respectively.
}
\label{picturereflectionequation}
\end{figure}

We also introduce the triangular $K$-matrix acting on
$W_a$ (see Figure \ref{picturekmatrix})
\begin{eqnarray}
K_{a}(u)=\left( 
\begin{array}{cc}
Bu-A & 0 \\
u-u^{-1} & Bu^{-1}-A \\
\end{array}
\right), \label{kmatrix}
\end{eqnarray}
where $A$ and $B$ are arbitrary complex parameters.
The matrix elements are explicitly given by
\begin{align}
{}_a \langle 0| K_{a}(u)
|0 \rangle_a&=Bu-A, \\
{}_a \langle 0| K_{a}(u)
|1 \rangle_a&=0, \\
{}_a \langle 1| K_{a}(u)
|1 \rangle_a&=Bu^{-1}-A, \\
{}_a \langle 1| K_{a}(u)
| 0 \rangle_a&=u-u^{-1}.
\end{align}

The $K$-matrix \eqref{kmatrix} together with the $R$-matrix
satisfy the following relation
\begin{align}
R_{ba}(u/w)K_b(u)R_{ab}(uw)K_a(w)
=K_a(w)R_{ba}(uw)K_b(u)R_{ab}(u/w),
\label{reflection equation}
\end{align}
which is called as the reflection equation
or the boundary Yang-Baxter equation \cite{Sklyanin}.
The reflection equation ensures the integrability at the boundary,
and we use this particular $K$-matrix
as local pieces of the wavefunctions at the boundary.
Note that \eqref{kmatrix} can be regarded as a specialization
of the full $K$-matrix which satisfies the reflection equation
\eqref{reflection equation}.
See Figure \ref{picturereflectionequation} for the grahical representation
of the reflection equation.

For later convenience,
we finally define in this section 
the following Pauli spin operators
$\sigma^+$ and $\sigma^-$ as operators acting on the (dual) orthonomal
basis as
\begin{align}
&\sigma^+|1 \rangle=|0 \rangle, \ 
\sigma^+|0 \rangle=0, \ 
\langle 0|\sigma^+=\langle 1|, \
\langle 1|\sigma^+=0, 
\\
&\sigma^-|0 \rangle=|1 \rangle, \
\sigma^-|1 \rangle=0, \
\langle 1|\sigma^-=\langle 0|, \
\langle 0|\sigma^-=0.
\end{align}

\begin{figure}[ht]
\includegraphics[width=12cm]{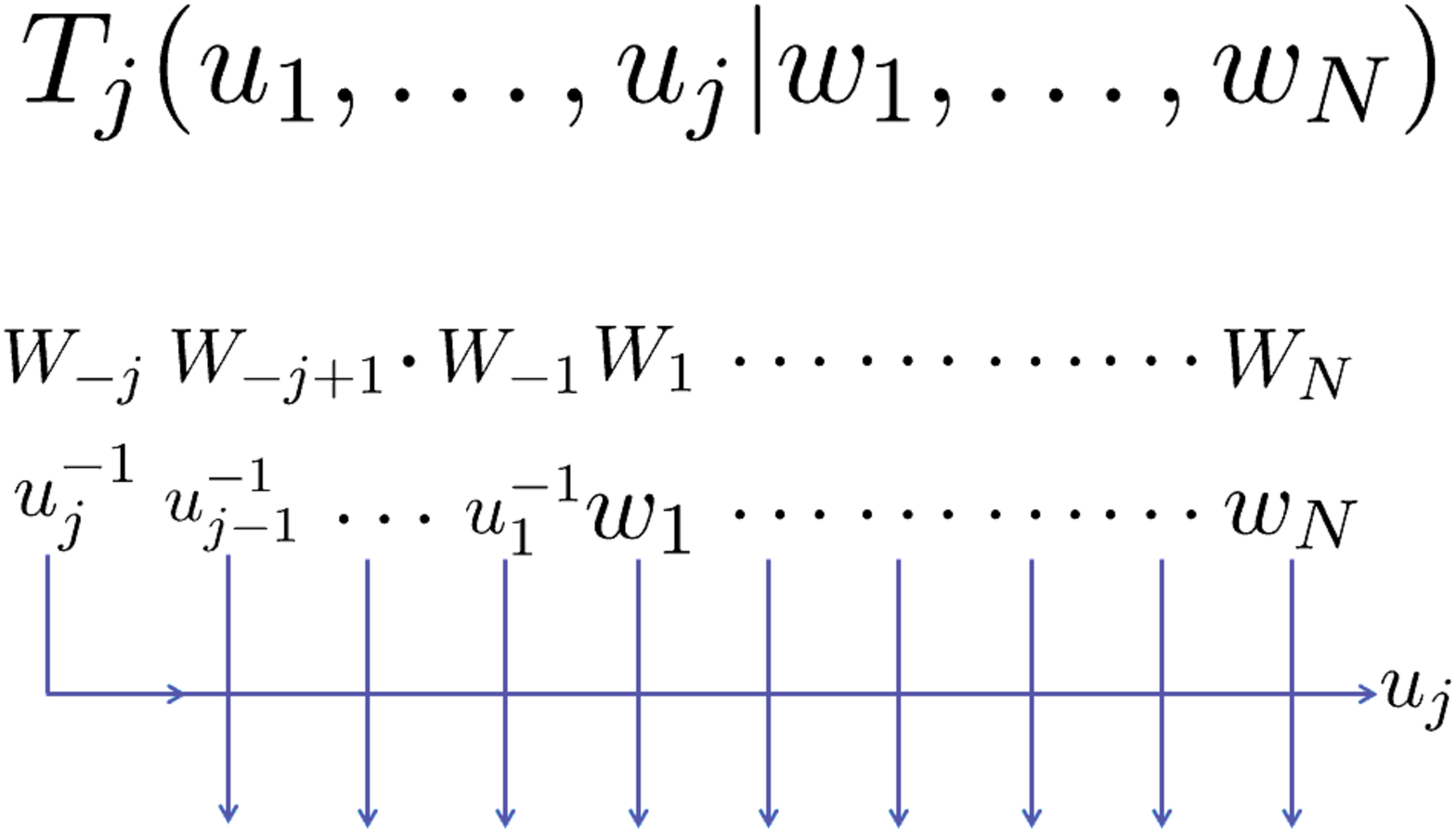}
\caption{The monodromy matrix $T_j(u_1,\dots,u_j|w_1,\dots,w_N)$
\eqref{monodromy1}.}
\label{picturemonodromy}
\end{figure}

\section{Wavefunctions}

To introduce the wavefunctions with
triangular boundary, we introduce the tensor product of the Fock spaces
$W_{-n} \otimes \cdots \otimes W_{-1} \otimes W_1 \otimes \cdots \otimes W_N$.
Note that here we do not introduce a Fock space named $W_0$.

We next define the following monodromy matrix
$T_j(u_1,\dots,u_j|w_1,\dots,w_N)$, $j=1,\dots,n$ using the $R$-matrix and the
$K$-matrix as
\begin{align}
&T_{j}(u_1,\dots,u_j|w_1,\dots,w_N) \nonumber \\
=&
R_{-j,N}(u_j,w_N)
\cdots
R_{-j,1}(u_j,w_1)
R_{-j,-1}(u_j u_1,1) \cdots
R_{-j,-j+1}(u_j u_{j-1},1)
K_{-j}(u_j),
\label{monodromy1}
\end{align}
which acts on $W_{-j} \otimes \cdots \otimes W_{-1}
\otimes W_1 \otimes \cdots \otimes W_N$.
See Figure \ref{picturemonodromy}
for a pictorial description of \eqref{monodromy1}.

Next, we define the state vector
$|\Psi_{N,n}(u_1,\dots,u_n|w_1,\dots,w_N) \rangle
\in W_1 \otimes \cdots \otimes W_N$
by using the monodromy matrix $T_j(u_1,\dots,u_j|w_1,\dots,w_N)$ as
\begin{align}
&|\Psi_{N,n}(u_1,\dots,u_n|w_1,\dots,w_N) \rangle \nonumber \\
=&\langle 0^n|
T_1(u_1|w_1,\dots,w_N) \cdots T_n(u_1,\dots,u_n|w_1,\dots,w_N)
|\Omega \rangle_{n+N},
\label{offshellBethevector}
\end{align}
where the states $\langle 0^n|$ and $|\Omega \rangle_{n+N}$
are defined as
\begin{align}
\langle 0^n|&={}_{-n} \langle 0| \otimes \cdots \otimes
{}_{-1} \langle 0|,
\\
|\Omega \rangle_{n+N}&=|0 \rangle_{-n} \otimes \cdots \otimes
|0 \rangle_{-1} \otimes |0 \rangle_1 \otimes \cdots \otimes
|0 \rangle_N.
\end{align}
This is an analogue of the off-shell Bethe vector.
We remark that there is a big difference on the properties
between the state vector
\eqref{offshellBethevector} and the ordinary off-shell Bethe vector.
The ordinary off-shell Bethe vector of the six-vertex model
(and its degeneration to the five-vertex model) is constructed from
the multiple action of the so-called $B$-operators on the vacuum state,
and one $B$-operator always creates one dowm spin,
hence the ordinary off-shell Bethe vector
constructed from $n$ layers of $B$-operators gives an $n$-down spin state.
However, the state vector \eqref{offshellBethevector}
constructed from $n$ layers
$T_j(u_1,\dots,u_j|w_1,\dots,w_N)$, $j=1,\cdots,n$
do not always give an $n$-down spin state.

Keeping this in mind,
let us now introduce an analogue of the wavefunctions
with triangular boundary.
We first introduce the following dual spin states
\begin{align}
\langle x_1 \cdots x_m|
&=(_1 \langle 0| \otimes \cdots \otimes {}_N \langle 0|)
\prod_{j=1}^m \sigma^+_{x_j}
\in W_1^* \otimes \cdots \otimes W_N^*
, \label{dualparticleconfiguration}
\end{align}
which are states labelling the configurations
of down spins
$1 \le x_1 < x_2 < \cdots < x_m \le N$.
$W_a^*$ means the dual space of $W_a$.

\begin{figure}[ht]
\includegraphics[width=12cm]{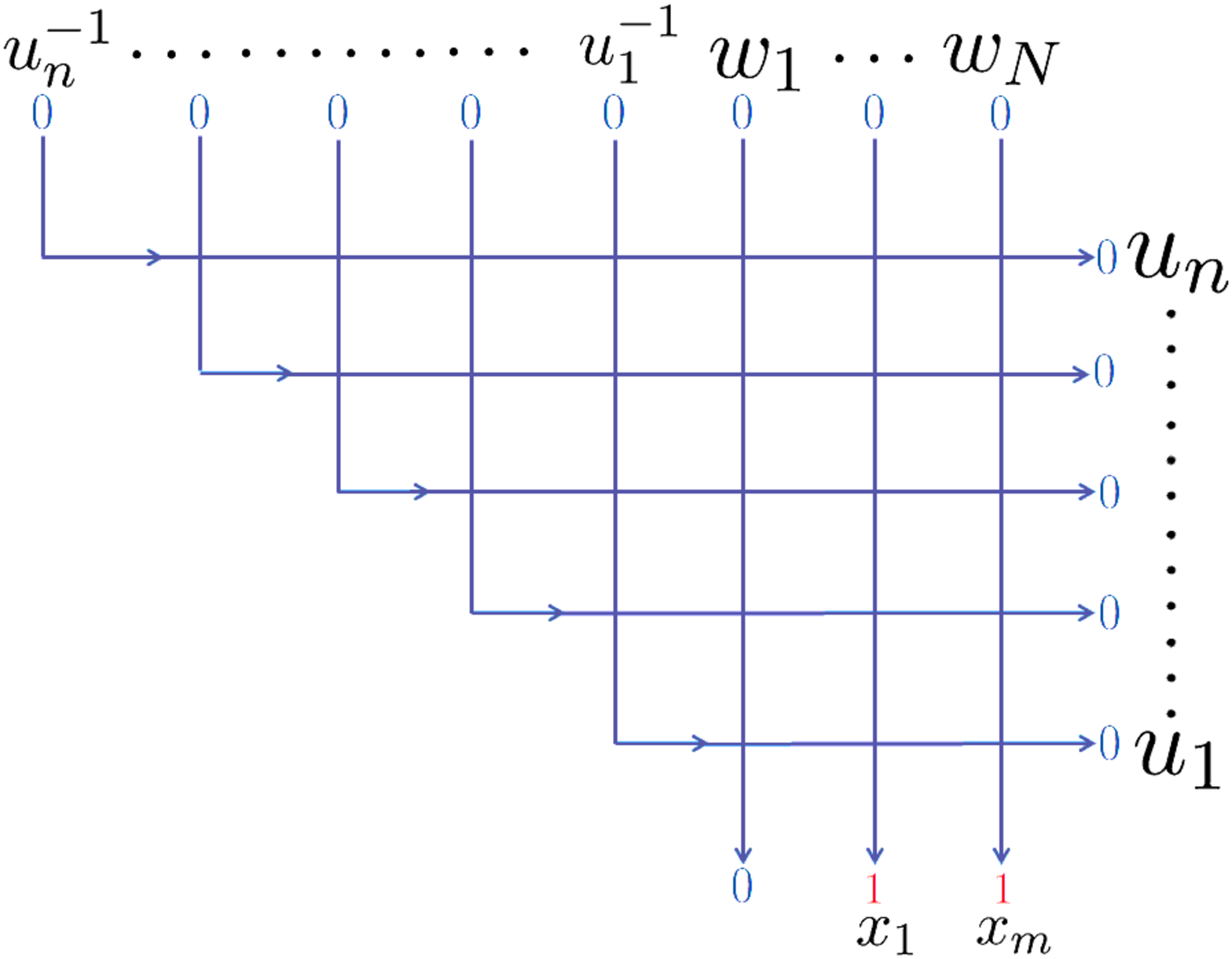}
\caption{The wavefunctions
with triangular boundary
$W_{N,n,m}(u_1,\dots,u_n|w_1,\dots,w_N|x_1,\dots,x_m)$
\eqref{DWBPF}.
This figure illustrates the case $N=3, \ n=5, \ m=2, \ x_1=2, \ x_2=3$.
}
\label{pictureDWBPF}
\end{figure}

Now we define the wavefunctions
with triangular boundary \\
$W_{N,n,m}(u_1,\dots,u_n|w_1,\dots,w_N|x_1,\dots,x_m)$
as the inner product between
the state vector \\
$|\Psi_{N,n}(u_1,\dots,u_n|w_1,\dots,w_N) \rangle$
and the $m$-down spin state $\langle x_1 \cdots x_m|$ as
\begin{align}
W_{N,n,m}(u_1,\dots,u_n|w_1,\dots,w_N|x_1,\dots,x_m)
=\langle x_1 \cdots x_m|\Psi_{N,n}(u_1,\dots,u_n|w_1,\dots,w_N) \rangle.
\label{wavefunction}
\end{align}

The wavefunctions \eqref{wavefunction}
can be seen as an extension of
the domain wall boundary partition function
with triangular boundary
\begin{align}
&Z_{n,m}(u_1,\dots,u_n|w_1,\dots,w_m) \nonumber \\
=&
\langle 0^n 1^m |
T_1(u_1|w_1,\dots,w_m) \cdots T_n(u_1,\dots,u_n|w_1,\dots,w_m)
|\Omega \rangle_{n+m}, \label{DWBPF} \\
\langle 0^n 1^m|=&{}_{-n} \langle 0| \otimes \cdots \otimes 
{}_{-1} \langle 0| \otimes
{}_{1} \langle 1| \otimes \cdots \otimes {}_{m} \langle 1|,
\end{align}
which we introduced in our last paper \cite{Motr},
since the special case $N=m$, $x_j=j$, $j=1,\cdots,m$ of the wavefunctions \eqref{wavefunction}
is nothing but the domain wall boundary partition function
\begin{align}
W_{m,n,m}(u_1,\dots,u_n|w_1,\dots,w_m|1,\dots,m)
=Z_{n,m}(u_1,\dots,u_n|w_1,\dots,w_m).
\end{align}

The wavefunctions
$W_{N,n,m}(u_1,\dots,u_n|w_1,\dots,w_N|x_1,\dots,x_m)$
is a symmetric function with respect to the spectral parameters
$\{ u \}_n$ which can be shown in the same way we showed
for the case of the domain wall boundary partition function
\cite{Motr} by the so-called railroad argument using the Yang-Baxter relation
and the reflection equation.
Thus, we sometimes
abbreviate $W_{N,n,m}(u_1,\dots,u_n|w_1,\dots,w_N|x_1,\dots,x_m)$ as \\
$W_{N,n,m}(\{ u \}_n|w_1,\dots,w_N|x_1,\dots,x_m)$ where
$\{ u \}_n$ means $\{ u_1,\dots,u_n \}$ as a set.
See Figure \ref{pictureDWBPF} for the graphical description of the
wavefunctions $W_{N,n,m}(u_1,\dots,u_n|w_1,\dots,w_N|x_1,\dots,x_m)$.

\begin{figure}[ht]
\includegraphics[width=12cm]{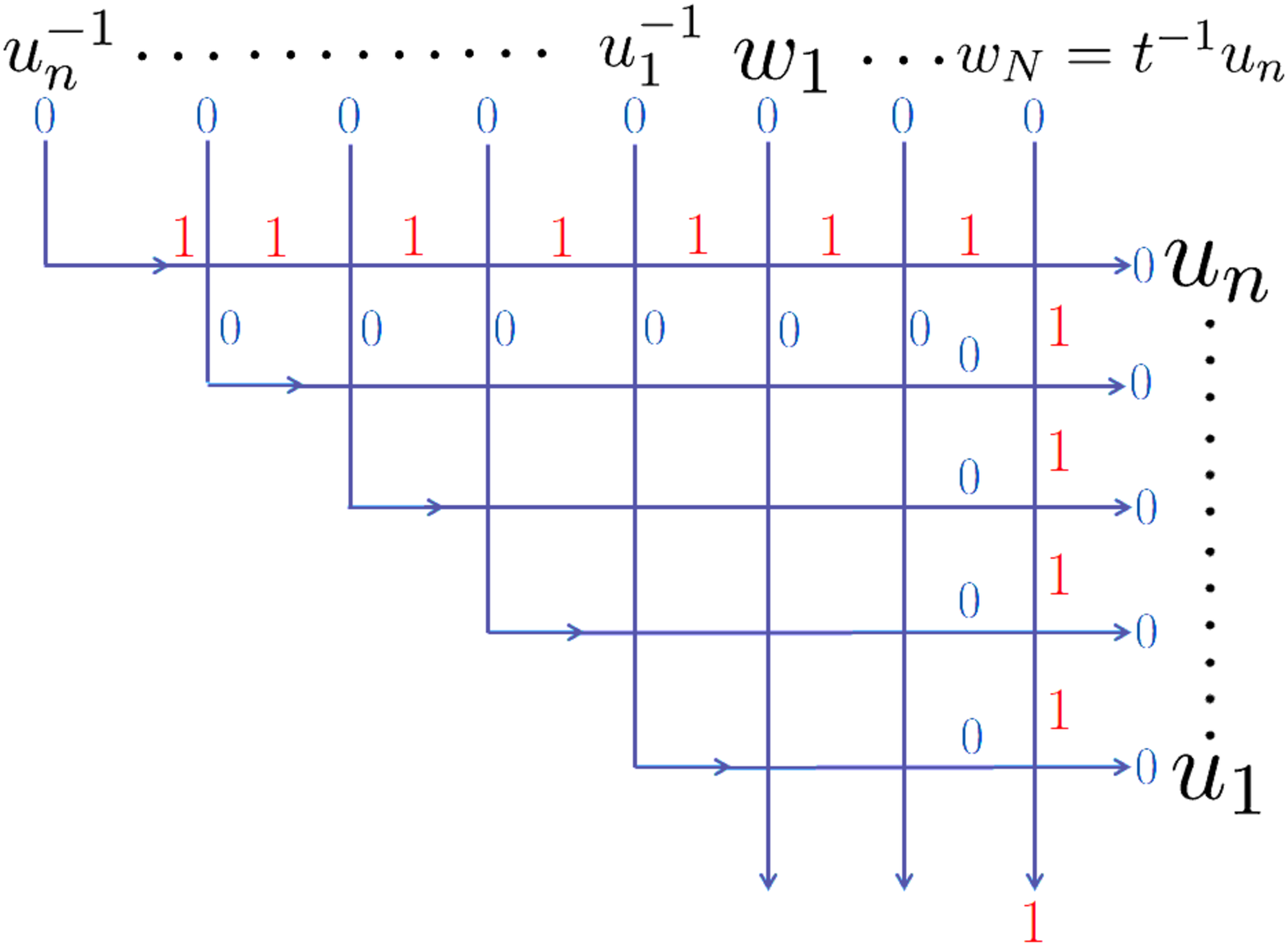}
\caption{The recursion relation
$W_{N,n,m}(u_1,\dots,u_n|w_1,\dots,w_N|x_1,\dots,x_m)$,
$x_m=N$ evaluated at $w_N=t^{-1}u_n$
\eqref{recursionwavefunction}
.}
\label{picturerecursion}
\end{figure}

\begin{figure}[ht]
\includegraphics[width=12cm]{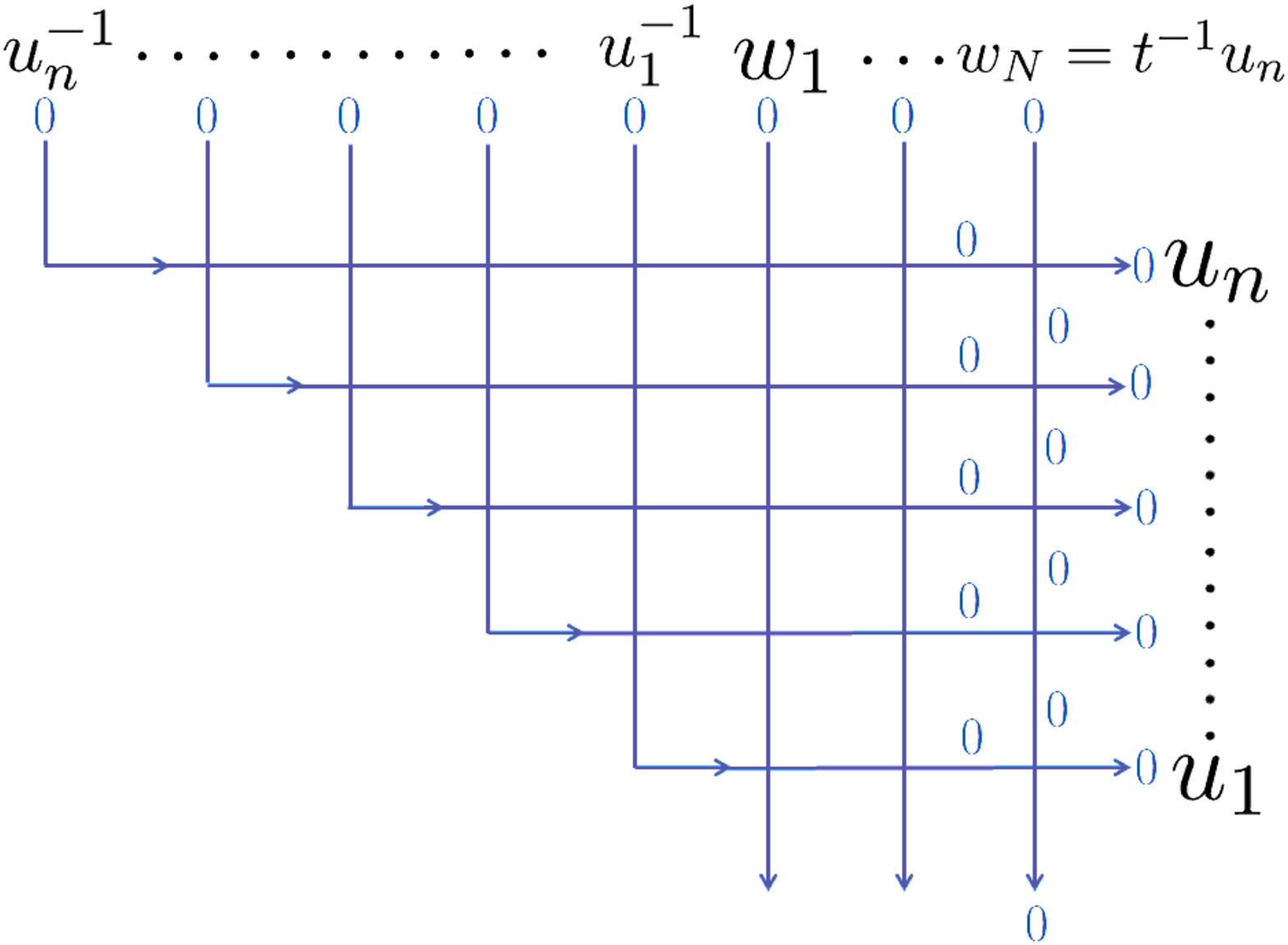}
\caption{The factorization of
$W_{N,n,m}(u_1,\dots,u_n|w_1,\dots,w_N|x_1,\dots,x_m)$,
$x_m \neq N$
\eqref{recursionwavefunction2}
.}
\label{picturerecursion2}
\end{figure}

\begin{figure}[ht]
\includegraphics[width=12cm]{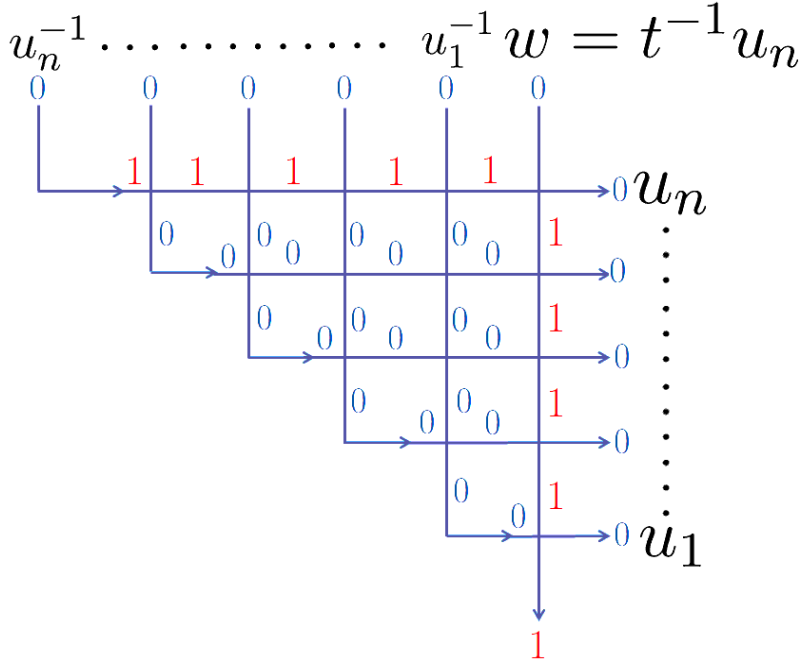}
\caption{The recursion relation
$W_{N,n,m}(u_1,\dots,u_n|w_1,\dots,w_N|x_1,\dots,x_m)$ evaluated at $w=t^{-1}u_n$
for the case $N=m=1$, $x_1=1$ \eqref{initialrecursion}.}
\label{pictureinitial}
\end{figure}

\begin{figure}[ht]
\includegraphics[width=12cm]{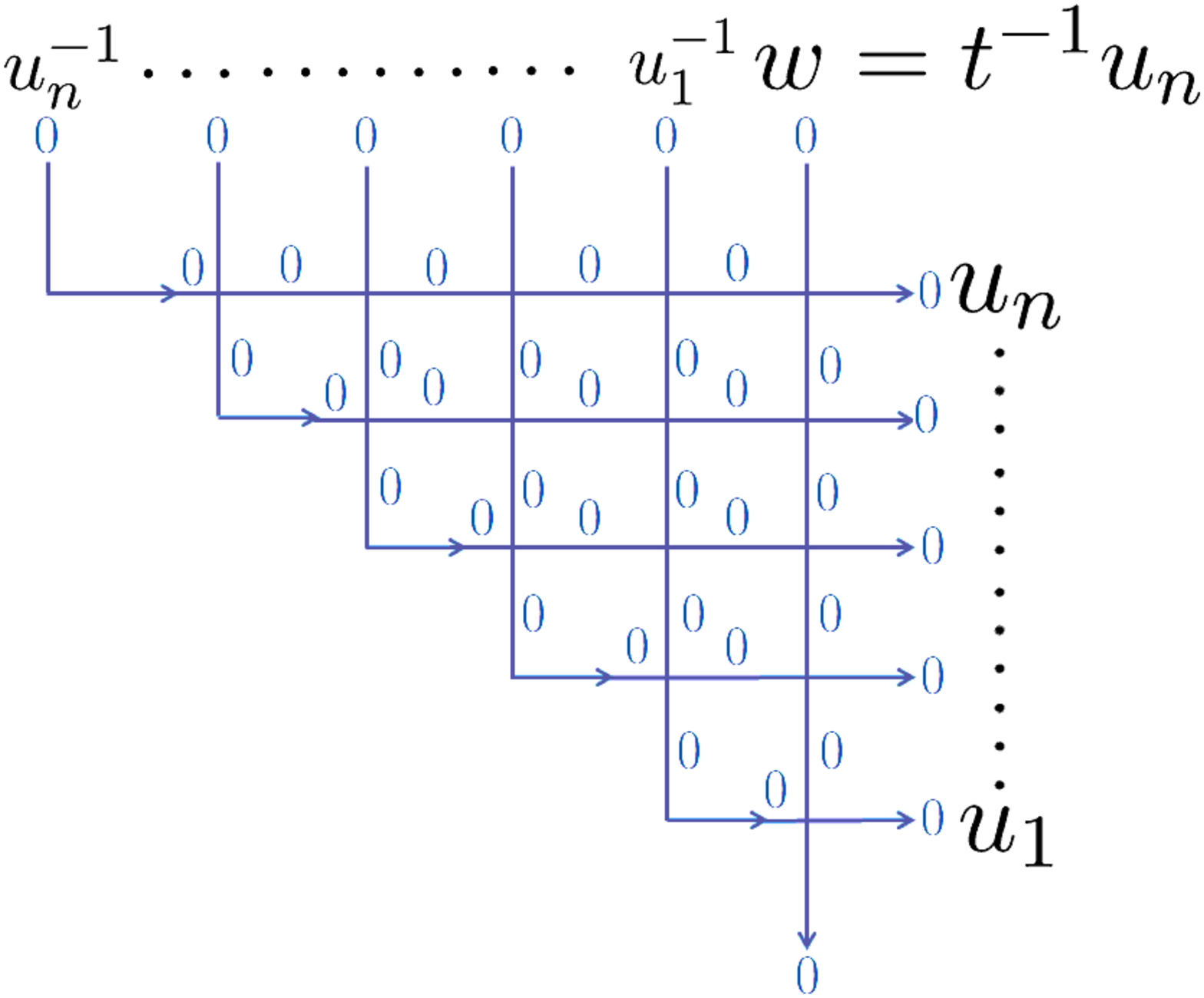}
\caption{The factorization of
$W_{N,n,m}(u_1,\dots,u_n|w_1,\dots,w_N|x_1,\dots,x_m)$
for the case $N=1, \ m=0$ \eqref{initialrecursion2}.}
\label{pictureinitial2}
\end{figure}

\section{Izergin-Korepin analysis}
In this section, we list and prove the properties
of the wavefunctions \\
$W_{N,n,m}(u_1,\dots,u_n|w_1,\dots,w_N|x_1,\dots,x_m)$
defined in the last section
by extending the Izergin-Korepin analysis \cite{Ko,Iz,Wheeler}
on the domain wall boundary partition function with triangular boundary
in our recent paper \cite{Motr}.

\begin{proposition} \label{propertiesfordomainwallboundarypartitionfunction}
The wavefunctions
$W_{N,n,m}(u_1,\dots,u_n|w_1,\dots,w_N|x_1,\dots,x_m)$
satisfies the following properties. \\
\\
 (1) $W_{N,n,m}(u_1,\dots,u_n|w_1,\dots,w_N|x_1,\dots,x_m)$
is a polynomial of degree $n-1$ in $w_N$
if $x_m=N$ and degree $n$ if $x_m \neq N$.
\\
 (2) $W_{N,n,m}(u_1,\dots,u_n|w_1,\dots,w_N|x_1,\dots,x_m)$ is symmetric
with respect to $u_j$, $j=1,\dots,n$.
\\
(3) The following recursive relations between the
wavefunctions hold if $x_m=N$
(Figure \ref{picturerecursion}):
\begin{align}
&W_{N,n,m}(u_1,\dots,u_n|w_1,\dots,w_N|x_1,\dots,x_m)|_{w_N=t^{-1}u_n}
\nonumber \\
=&(1-t) \prod_{j=1}^{n-1}(tu_j-u_n)
\prod_{j=1}^{n}(u_j u_n-1) \prod_{j=1}^{N-1}(u_n-w_j)
\nonumber \\
&\times W_{N-1,n-1,m-1}(u_1,\dots,u_{n-1}|w_1,\dots,w_{N-1}|x_1,\dots,x_{m-1})
. \label{recursionwavefunction}
\end{align}
If $x_m \neq N$, the following factorizations hold for the wavefunctions
(Figure \ref{picturerecursion2}):
\begin{align}
&W_{N,n,m}(u_1,\dots,u_n|w_1,\dots,w_N|x_1,\dots,x_m) \nonumber \\
=&\prod_{j=1}^n (u_j-tw_N)
W_{N-1,n,m}(u_1,\dots,u_n|w_1,\dots,w_{N-1}|x_1,\dots,x_m).
\label{recursionwavefunction2}
\end{align}
\\
(4) The following explicit evaluations hold for the case $N=m=1$
(Figure \ref{pictureinitial}):
\begin{align}
&
W_{1,n,1}(u_1,\dots,u_n|w|1)|_{w=t^{-1}u_n}
\nonumber \\
=&(1-t) \prod_{j=1}^{n-1}(tu_j-u_n)
(Bu_j-A)
\prod_{j=1}^{n}(u_j u_n-1)
\prod_{1 \le j < k \le n-1}(u_ju_k-t).
\label{initialrecursion}
\end{align}
If $N=1, \ m=0$, the following explicit evaluation holds
(Figure \ref{pictureinitial2}):
\begin{align}
W_{1,n,0}(u_1,\dots,u_n|w)=\prod_{j=1}^n (Bu_j-A)
\prod_{j=1}^n (u_j-tw) \prod_{1 \le j < k \le n}(u_j u_k-t).
\label{initialrecursion2}
\end{align}
\end{proposition}

\begin{proof}
Property (1) for the case $x_m=N$ can be easily shown in the same way with
the domain wall boundary partition function (with triangular boundary)
by inserting the completeness relation
in one spin down state sector.

For the case $x_m \neq N$, one can also easily show by
its graphical representation and
the ice rule of the $R$-matrix
$
{}_a \langle \gamma | {}_b \langle \delta | R_{a b}(u,w)
|\alpha \rangle_a | \beta \rangle_b=0$
unless $\alpha+\beta=\gamma+\delta$.
Using the ice rule, one finds that the $R$-matrices on the rightmost row
freeze (Figure \ref{picturerecursion2}).
The contribution to the weights from the freezed rightmost row is
$\prod_{j=1}^n (u_j-tw_N)$,
and the remaining part is nothing but the wavefunctions
$W_{N-1,n,m}(u_1,\dots,u_n|w_1,\dots,w_{N-1}|x_1,\dots,x_m)$.
This shows that $W_{N,n,m}(u_1,\dots,u_n|w_1,\dots,w_N|x_1,\dots,x_m)$
has the following factorization for the case $x_m \neq N$
\begin{align}
&W_{N,n,m}(u_1,\dots,u_n|w_1,\dots,w_N|x_1,\dots,x_m) \nonumber \\
=&\prod_{j=1}^n (u_j-tw_N)
W_{N-1,n,m}(u_1,\dots,u_n|w_1,\dots,w_{N-1}|x_1,\dots,x_m),
\end{align}
which is nothing but
\eqref{recursionwavefunction2}.
This relation shows
Properties (1) and (3) for the case $x_m \neq N$.

Property (2) can be shown by the standard railroad argument
using the Yang-Baxter relation and the reflection equation repeatedly,
exactly in the same way we proved for the case of the
domain wall boundary partition function \cite{Motr}.
We remark that the wavefunctions
$W_{N,n,m}(u_1,\dots,u_n|w_1,\dots,w_N|x_1,\dots,x_m)$
is no more symmetric with respect to $w_j$, $j=1,\dots,m$
in general unlike the domain wall boundary partition function.
This comes from the fact that the dual spin state
$\langle x_1 \cdots x_m|$ which is used to construct
the wavefunctions is not a consecutive sequence
of down spins anymore in general.
Note that the dual spin state
$\langle x_1 \cdots x_m|$ does not have any influence on
the railroad argument to prove the commutativity of
the spectral parameters $\{ u \}_n$.

Property (3) for the case $x_m=N$
can be proved with the help of the graphical representation
of the wavefunctions $W_{N,n,m}(u_1,\dots,u_n|w_1,\dots,w_N|x_1,\dots,x_m)$,
and can be showed essentially in the same way with proving
a similiar property for
the domain wall boundary partition function
$Z_{n,m}(\{ u \}_n|\{ w \}_m)$ \cite{Motr}.
To show \eqref{recursionwavefunction},
let us see what happens when one sets $w_N$ as $w_N=t^{-1}u_n$
(see Figure \ref{picturerecursion}).
One first sees that the $R$-matrix at the top right corner freezes
due to the vanishing property
${}_{-n} \langle 0 | {}_{N} \langle 0 | R(u_n,w_N=t^{-1}u_n)|0 \rangle_{-n} |0 \rangle_{N}=0$.
Then using the ice rule of the $R$-matrix,
one finds the top row and the rightmost column freeze.
The total contribution of the weights from the freezed part
can be calculated by multiplying all the matrix elements of
the $R$-matrix and the $K$-matrix which appear,
and we find it is given by
$(1-t)\prod_{j=1}^{n-1}(tu_j-u_n) \prod_{j=1}^n(u_j u_n-1)
\prod_{j=1}^{N-1}(u_n-w_j)$.
Next, one finds that the remaining unfreezed part
is nothing but the wavefunctions
$W_{N-1,n-1,m-1}(u_1,\dots,u_{n-1}|w_1,\dots,w_{N-1}|x_1,\dots,x_{m-1})$
,
from which one concludes that
$W_{N,n,m}(u_1,\dots,u_n|w_1,\dots,w_N|x_1,\dots,x_m)$
evaluated at $w_N=t^{-1}u_n$
is the product of
$(1-t)\prod_{j=1}^{n-1}(tu_j-u_n) \prod_{j=1}^n(u_j u_n-1)
\prod_{j=1}^{N-1}(u_n-w_j)$
and \\
$W_{N-1,n-1,m-1}(u_1,\dots,u_{n-1}|w_1,\dots,w_{N-1}|x_1,\dots,x_{m-1})$.
This shows \eqref{recursionwavefunction}.

What remains to be shown is
Property (4), which can be regarded as the initial condition
of the recursion relation (Property (3)).

Property (4) for the case $m=1$ is shown in
\cite{Motr}.
We include the proof here for the sake of completeness.
This can be shown in the same way as
Property (3).
Let us denote $w_1$ as $w$ for simplicity.
From its graphical representation,
one finds great simplication occurs
for the wavefunction
$W_{1,n,1}(u_1,\dots,u_n|w|1)$
(Figure \ref{pictureinitial}).
When one sets $w=t^{-1}u_n$,
one first finds that the top row and the rightmost column
freeze.
We make a further observation on $W_{1,n,1}(u_1,\dots,u_n|w|1)$.
Using the property ${}_{a} \langle 0|K_a(u)|1 \rangle_a=0$
of the triangular $K$-matrix \eqref{kmatrix},
one finds that all spins on the bottom row freeze,
and this freezing process continues and we finally find
that all spins get freezed.
Multiplying all the appearing matrix elements of
the $R$-matrix and the $K$-matrix,
one finds the total contribution is given by
$(1-t) \prod_{j=1}^{n-1}(tu_j-u_n)
(Bu_j-A)
\prod_{j=1}^{n}(u_j u_n-1)
\prod_{1 \le j < k \le n-1}(u_ju_k-t)$,
hence \eqref{initialrecursion} is shown.

Property (4) for the case $m=0$ can also be proved in the same
way. By drawing the graphical representation
of $W_{1,n,0}(u_1,\dots,u_n|w)$,
one finds by using the ice rule of the $R$-matrix
and the property ${}_{a} \langle 0|K_a(u)|1 \rangle_a=0$
of the $K$-matrix that all spins get freezed.
Multiplying all the appearing matrix elements of
the $R$-matrix and the $K$-matrix,
one finds the total contribution to the wavefunctions is given by
$\prod_{j=1}^n (Bu_j-A)
\prod_{j=1}^n (u_j-tw) \prod_{1 \le j < k \le n}(u_j u_k-t)$,
from which one concludes \eqref{initialrecursion2} holds.
See Figure \ref{pictureinitial2} for the graphical representation
of $W_{1,n,0}(u_1,\dots,u_n|w)$.

\end{proof}

\section{Symmetric functions}
In this section, we introduce
a class of symmetric function,
and prove that the symmetric function represents 
the wavefunctions of the six-vertex model with triangular boundary
$W_{N,n,m}(u_1,\dots,u_n|w_1,\dots,w_N|x_1,\dots,x_m)$
by showing that it satisfies all the required properties derived
by the Izergin-Korepin analysis in the last section.

\begin{definition}
We define the following symmetric function \\
$F_{N,n,m}(u_1,\dots,u_n|w_1,\dots,w_N|x_1,\dots,x_m)$
which depends on the symmetric variables $u_1,\dots,u_n$,
complex parameters $w_1,\dots,w_N$
and integers $x_1,\dots,x_m$ satisfying
$1 \le x_1 < \cdots < x_m \le N$,
\begin{align}
&F_{N,n,m}(u_1,\dots,u_n|w_1,\dots,w_N|x_1,\dots,x_m) \nonumber \\
=
&\frac{1}{(n-m)!}
\sum_{\sigma \in S_n}
\prod_{j=1}^m \prod_{k=x_j+1}^N (u_{\sigma(j)}-tw_k)
\prod_{1 \le j < k \le m}
\frac{tu_{\sigma(j)}-u_{\sigma(k)}}{u_{\sigma(j)}-u_{\sigma(k)}}
(u_{\sigma(j)}u_{\sigma(k)}-1) \nonumber \\
&\times
\prod_{\substack{m+1 \le j \le n \\ 1 \le k \le N}}(u_{\sigma(j)}-tw_k)
\prod_{\substack{m+1 \le j \le n \\ 1 \le k \le m}}
\frac{tu_{\sigma(j)}-u_{\sigma(k)}}{u_{\sigma(j)}-u_{\sigma(k)}}
(u_{\sigma(j)}u_{\sigma(k)}-1) \nonumber \\
&\times
\prod_{j=1}^m \prod_{k=1}^{x_j-1}(u_{\sigma(j)}-w_k)
\prod_{m+1 \le j < k \le n}(u_{\sigma(j)}u_{\sigma(k)}-t)
\nonumber \\
&\times
\prod_{j=1}^m (1-t)(u_{\sigma(j)}^2-1)
\prod_{j=m+1}^n (Bu_{\sigma(j)}-A).
\label{righthandside}
\end{align}
\end{definition}

\begin{theorem}
The
wavefunctions of the six-vertex model with triangular boundary
\\
$Z_{N,n,m}(u_1,\dots,u_n|w_1,\dots,w_N|x_1,\dots,x_m)$
is explicitly expressed as the
symmetric function
$F_{N,n,m}(u_1,\dots,u_n|w_1,\dots,w_N|x_1,\dots,x_m)$
\begin{align}
Z_{N,n,m}(u_1,\dots,u_n|w_1,\dots,w_N|x_1,\dots,x_m)=
F_{N,n,m}(u_1,\dots,u_n|w_1,\dots,w_N|x_1,\dots,x_m).
\end{align}
\end{theorem}

\begin{proof}
We prove this theorem by showing that
the symmetric funcion
\eqref{righthandside} \\
$F_{N,n,m}(u_1,\dots,u_n|w_1,\dots,w_N|x_1,\dots,x_m)$
satisfies all the four Properties in
Proposition \ref{propertiesfordomainwallboundarypartitionfunction}.

Let us first show Property (1).
First, note that the factor
$\displaystyle \prod_{j=1}^m \prod_{k=x_j+1}^N (u_{\sigma(j)}-tw_k)$
in
$F_{N,n,m}(u_1,\dots,u_n|w_1,\dots,w_N|x_1,\dots,x_m)$
is a polynomial of degree $m-1$ in $w_N$ if $x_m=N$
and degree $m$ if $x_m \neq N$.
There is also a factor
$\displaystyle \prod_{m+1 \le j \le n}(u_{\sigma(j)}-tw_N)$
in \eqref{righthandside}
which contributes to the degree of $w_N$,
and one finds that
$F_{N,n,m}(u_1,\dots,u_n|w_1,\dots,w_N|x_1,\dots,x_m)$
is a polynomial of degree $n-1$ in $w_N$ if $x_m=N$
and degree $n$ if $x_m \neq N$.

It is also easy to find that
$F_{N,n,m}(u_1,\dots,u_n|w_1,\dots,w_N|x_1,\dots,x_m)$
is symmetric with respect to $u_j$, $j=1,\dots,n$,
since the sum is over all permutations
of the variables $u_j$, $j=1,\dots,n$.

Let us show Property (3).
We first show the function
$F_{N,n,m}(\{ u \}_n|w_1,\dots,w_N|x_1,\dots,x_m)$ satisfies
\eqref{recursionwavefunction}
which we have to prove for the case $x_m=N$.
In this case, first note that the factor
\begin{align}
\prod_{j=1}^m \prod_{k=x_j+1}^N(u_{\sigma(j)}-tw_k)
\prod_{\substack{m+1 \le j \le n \\ 1 \le k \le N}}(u_{\sigma(j)}-tw_k),
\end{align}
in each summand essentially becomes
\begin{align}
\prod_{j=1}^{m-1} \prod_{k=x_j+1}^N(u_{\sigma(j)}-tw_k)
\prod_{\substack{m+1 \le j \le n \\ 1 \le k \le N}}(u_{\sigma(j)}-tw_k).
\label{factorconsideration}
\end{align}
Then one can extract a factor
$\displaystyle
\prod_{j=1}^{m-1} (u_{\sigma(j)}-tw_N)
\prod_{j=m+1}^N (u_{\sigma(j)}-tw_N)
$ from \eqref{factorconsideration}.
Let us concentrate on this factor.
If one substitutes $w_N=t^{-1}u_n$, 
this factor vanishes unless $\sigma$ satisfies $\sigma(m)=n$.

Therefore, only the summands satisfying $\sigma(m)=n$ 
in \eqref{righthandside} survive
after the substitution $w_N=t^{-1}u_n$.
Keeping this in mind, one rewrites
$F_{N,n,m}(\{ u \}_n|w_1,\dots,w_N|x_1,\dots,x_m)|_{w_N=t^{-1}u_n}$
by using the symmetric group $S_{n-1}$
where every $\sigma^\prime \in S_{n-1}$ satisfies
$\{\sigma^\prime(1),\cdots,\sigma^\prime(m-1),\sigma^\prime(m+1),\cdots,\sigma^\prime(n)\}=\{1,\cdots,n-1 \}$ as follows:
\begin{align}
&F_{N,n,m}(u_1,\dots,u_n|w_1,\dots,w_N|x_1,\dots,x_m)
|_{w_N=t^{-1}u_n} \nonumber \\
=&\frac{1}{(n-m)!}
\sum_{\sigma^\prime \in S_{n-1}}
\prod_{j=1}^{m-1} \prod_{k=x_j+1}^{N-1}(u_{\sigma^\prime(j)}-tw_k)
\prod_{1 \le j < k \le m-1}
\frac{tu_{\sigma^\prime(j)}-u_{\sigma^\prime(k)}}{u_{\sigma^\prime(j)}
-u_{\sigma^\prime(k)}}
(u_{\sigma^\prime(j)}u_{\sigma^\prime(k)}-1) \nonumber \\
&\times
\prod_{\substack{m+1 \le j \le n \\ 1 \le k \le N-1}}
(u_{\sigma^\prime(j)}-tw_k)
\prod_{\substack{m+1 \le j \le n \\ 1 \le k \le m-1}}
\frac{tu_{\sigma^\prime(j)}-u_{\sigma^\prime(k)}}{u_{\sigma^\prime(j)}
-u_{\sigma^\prime(k)}}
(u_{\sigma^\prime(j)}u_{\sigma^\prime(k)}-1) \nonumber \\
&\times \prod_{j=1}^{m-1}(tu_{\sigma^\prime(j)}-u_n)(u_{\sigma^\prime(j)}u_n-1)
\prod_{j=m+1}^{n}(tu_{\sigma^\prime(j)}-u_n)(u_{\sigma^\prime(j)}u_n-1)
\nonumber \\
&\times
\prod_{j=1}^{m-1} \prod_{k=1}^{x_j-1}(u_{\sigma^\prime(j)}-w_k)
\prod_{k=1}^{N-1}(u_n-w_k)
\prod_{m+1 \le j < k \le n}(u_{\sigma^\prime(j)}u_{\sigma^\prime(k)}-t)
\nonumber \\
&\times
(1-t)(u_n^2-1)
\prod_{j=1}^{m-1} (1-t)(u_{\sigma^\prime(j)}^2-1)
\prod_{j=m+1}^n (Bu_{\sigma^\prime(j)}-A).
\label{righthandsideaftersubstitution}
\end{align}
One easily notes that
the factors $\displaystyle
\prod_{k=1}^{N-1}(u_n-w_k)$
and $(1-t)(u_n^2-1)$ in the sum are independent of the permutation
$S^\prime_{n-1}$.
One also finds the factor
$\displaystyle \prod_{j=1}^{m-1}(tu_{\sigma^\prime(j)}-u_n)(u_{\sigma^\prime(j)}u_n-1)
\prod_{j=m+1}^{n}(tu_{\sigma^\prime(j)}-u_n)(u_{\sigma^\prime(j)}u_n-1)$
is also independent of $S^\prime_{n-1}$ since we have
\begin{align}
&\prod_{j=1}^{m-1}(tu_{\sigma^\prime(j)}-u_n)(u_{\sigma^\prime(j)}u_n-1)
\prod_{j=m+1}^{n}(tu_{\sigma^\prime(j)}-u_n)(u_{\sigma^\prime(j)}u_n-1)
\nonumber \\
=&\prod_{j=1}^{n-1}(tu_j-u_n)(u_j u_n-1).
\end{align}
Thus, \eqref{righthandsideaftersubstitution} can be rewritten furthermore as
\begin{align}
&F_{N,n,m}(u_1,\dots,u_n|w_1,\dots,w_N|x_1,\dots,x_m)|_{w_N=t^{-1}u_n} \nonumber \\
=&
(1-t)(u_n^2-1) \prod_{j=1}^{n-1}(tu_j-u_n)(u_j u_n-1)
\prod_{k=1}^{N-1}(u_n-w_k) \nonumber \\
&\times \frac{1}{(n-m)!}
\sum_{\sigma^\prime \in S_{n-1}}
\prod_{j=1}^{m-1} \prod_{k=x_j+1}^{N-1} (u_{\sigma^\prime(j)}-tw_k)
\prod_{1 \le j < k \le m-1}
\frac{tu_{\sigma^\prime(j)}-u_{\sigma^\prime(k)}}{u_{\sigma^\prime(j)}
-u_{\sigma^\prime(k)}}
(u_{\sigma^\prime(j)}u_{\sigma^\prime(k)}-1) \nonumber \\
&\times
\prod_{\substack{m+1 \le j \le n \\ 1 \le k \le N-1}}
(u_{\sigma^\prime(j)}-tw_k)
\prod_{\substack{m+1 \le j \le n \\ 1 \le k \le m-1}}
\frac{tu_{\sigma^\prime(j)}-u_{\sigma^\prime(k)}}{u_{\sigma^\prime(j)}
-u_{\sigma^\prime(k)}}
(u_{\sigma^\prime(j)}u_{\sigma^\prime(k)}-1) \nonumber \\
&\times
\prod_{j=1}^{m-1} \prod_{k=1}^{x_j-1}(u_{\sigma^\prime(j)}-w_k)
\prod_{m+1 \le j < k \le n}(u_{\sigma^\prime(j)}u_{\sigma^\prime(k)}-t)
\nonumber \\
&\times
\prod_{j=1}^{m-1} (1-t)(u_{\sigma^\prime(j)}^2-1)
\prod_{j=m+1}^n (Bu_{\sigma^\prime(j)}-A).
\label{righthandsideaftersubstitutiontwo}
\end{align}
Noting
\begin{align}
&\frac{1}{(n-m)!}
\sum_{\sigma^\prime \in S_{n-1}}
\prod_{j=1}^{m-1} \prod_{k=x_j+1}^{N-1} (u_{\sigma^\prime(j)}-tw_k)
\prod_{1 \le j < k \le m-1}
\frac{tu_{\sigma^\prime(j)}-u_{\sigma^\prime(k)}}{u_{\sigma^\prime(j)}
-u_{\sigma^\prime(k)}}
(u_{\sigma^\prime(j)}u_{\sigma^\prime(k)}-1) \nonumber \\
&\times
\prod_{\substack{m+1 \le j \le n \\ 1 \le k \le N-1}}
(u_{\sigma^\prime(j)}-tw_k)
\prod_{\substack{m+1 \le j \le n \\ 1 \le k \le m-1}}
\frac{tu_{\sigma^\prime(j)}-u_{\sigma^\prime(k)}}{u_{\sigma^\prime(j)}
-u_{\sigma^\prime(k)}}
(u_{\sigma^\prime(j)}u_{\sigma^\prime(k)}-1) \nonumber \\
&\times
\prod_{j=1}^{m-1} \prod_{k=1}^{x_j-1}(u_{\sigma^\prime(j)}-w_k)
\prod_{m+1 \le j < k \le n}(u_{\sigma^\prime(j)}u_{\sigma^\prime(k)}-t)
\nonumber \\
&\times
\prod_{j=1}^{m-1} (1-t)(u_{\sigma^\prime(j)}^2-1)
\prod_{j=m+1}^n (Bu_{\sigma^\prime(j)}-A) \nonumber \\
=&
F_{N-1,n-1,m-1}(u_1,\dots,u_{n-1}|w_1,\dots,w_{N-1}|x_1,\dots,x_{m-1})
,
\end{align}
one finds that
\eqref{righthandsideaftersubstitutiontwo} is nothing but the
following recursion relation
for the symmetric function
$F_{N,n,m}(u_1,\dots,u_n|w_1,\dots,w_N|x_1,\dots,x_m)$
\begin{align}
&F_{N,n,m}(u_1,\dots,u_n|w_1,\dots,w_N|x_1,\dots,x_m)|_{w_N=t^{-1}u_n} \nonumber \\
=&(1-t)\prod_{j=1}^{n-1}(tu_j-u_n)\prod_{j=1}^{n}(u_j u_n-1)
\prod_{j=1}^{N-1}(u_n-w_j) \nonumber \\
&\times 
F_{N-1,n-1,m-1}(u_1,\dots,u_{n-1}|w_1,\dots,w_{N-1}|x_1,\dots,x_{m-1}),
\end{align}
which is exactly the same recursion relation the wavefunctions
\\
$W_{N,n,m}(u_1,\dots,u_n|w_1,\dots,w_N|x_1,\dots,x_m)$ must satisfy.
Hence, Property (3) for the case $x_m=N$ is proved.

The case $x_m \neq N$ can be shown in a similar but much simpler way.
In this case we can rewrite $F_{N,n,m}(u_1,\dots,u_n|w_1,\dots,w_N|x_1,\dots,x_m)$ as
\begin{align}
&F_{N,n,m}(u_1,\dots,u_n|w_1,\dots,w_N|x_1,\dots,x_m) \nonumber \\
=
&\frac{1}{(n-m)!}
\sum_{\sigma \in S_n}
\prod_{j=1}^m \prod_{k=x_j+1}^{N-1} (u_{\sigma(j)}-tw_k)
\prod_{j=1}^m (u_{\sigma(j)}-tw_N)
\prod_{1 \le j < k \le m}
\frac{tu_{\sigma(j)}-u_{\sigma(k)}}{u_{\sigma(j)}-u_{\sigma(k)}}
(u_{\sigma(j)}u_{\sigma(k)}-1) \nonumber \\
&\times
\prod_{\substack{m+1 \le j \le n \\ 1 \le k \le N-1}}(u_{\sigma(j)}-tw_k)
\prod_{j=m+1}^n (u_{\sigma(j)}-tw_N)
\prod_{\substack{m+1 \le j \le n \\ 1 \le k \le m}}
\frac{tu_{\sigma(j)}-u_{\sigma(k)}}{u_{\sigma(j)}-u_{\sigma(k)}}
(u_{\sigma(j)}u_{\sigma(k)}-1) \nonumber \\
&\times
\prod_{j=1}^m \prod_{k=1}^{x_j-1}(u_{\sigma(j)}-w_k)
\prod_{m+1 \le j < k \le n}(u_{\sigma(j)}u_{\sigma(k)}-t)
\nonumber \\
&\times
\prod_{j=1}^m (1-t)(u_{\sigma(j)}^2-1)
\prod_{j=m+1}^n (Bu_{\sigma(j)}-A).
\label{righthandsidenew}
\end{align}
The product of the factors
$\displaystyle \prod_{j=1}^m (u_{\sigma(j)}-tw_N)$ and
$\displaystyle \prod_{j=m+1}^n (u_{\sigma(j)}-tw_N)$
becomes
\begin{align}
\displaystyle \prod_{j=1}^m (u_{\sigma(j)}-tw_N)
\displaystyle \prod_{j=m+1}^n (u_{\sigma(j)}-tw_N)
=\prod_{j=1}^n (u_j-tw_N),
\end{align}
which becomes independent of the permutation $S_n$.
Taking this into account, \eqref{righthandsidenew}
can be rewritten as
\begin{align}
&F_{N,n,m}(u_1,\dots,u_n|w_1,\dots,w_N|x_1,\dots,x_m) \nonumber \\
=&\prod_{j=1}^n (u_j-tw_N)
\frac{1}{(n-m)!}
\sum_{\sigma \in S_n}
\prod_{j=1}^m \prod_{k=x_j+1}^{N-1} (u_{\sigma(j)}-tw_k)
\prod_{1 \le j < k \le m}
\frac{tu_{\sigma(j)}-u_{\sigma(k)}}{u_{\sigma(j)}-u_{\sigma(k)}}
(u_{\sigma(j)}u_{\sigma(k)}-1) \nonumber \\
&\times
\prod_{\substack{m+1 \le j \le n \\ 1 \le k \le N-1}}(u_{\sigma(j)}-tw_k)
\prod_{\substack{m+1 \le j \le n \\ 1 \le k \le m}}
\frac{tu_{\sigma(j)}-u_{\sigma(k)}}{u_{\sigma(j)}-u_{\sigma(k)}}
(u_{\sigma(j)}u_{\sigma(k)}-1) \nonumber \\
&\times
\prod_{j=1}^m \prod_{k=1}^{x_j-1}(u_{\sigma(j)}-w_k)
\prod_{m+1 \le j < k \le n}(u_{\sigma(j)}u_{\sigma(k)}-t)
\nonumber \\
&\times
\prod_{j=1}^m (1-t)(u_{\sigma(j)}^2-1)
\prod_{j=m+1}^n (Bu_{\sigma(j)}-A)
\nonumber \\
=&\prod_{j=1}^n (u_j-tw_N)
F_{N-1,n,m}(u_1,\dots,u_n|w_1,\dots,w_{N-1}|x_1,\dots,x_m),
\end{align}
which is also exactly the recursion relation the wavefunctions \\
$W_{N,n,m}(u_1,\dots,u_n|w_1,\dots,w_N|x_1,\dots,x_m)$
must satisfy for the case $x_m \neq N$.

Let us finally show Property (4).
The case $m=1$ is proved in \cite{Motr}.
We display the proof here for completeness.
This can be shown by making further analysis on
the symmetric function
$F_{N,n,m}(\{ u \}_n|w_1,\dots,w_N|x_1,\dots,x_m)
|_{w_N=t^{-1}u_n}$
for the case $N=m=1$, $x_1=1$. We first write down the case $N=m=1$, $x_1=1$
of \eqref{righthandsideaftersubstitutiontwo}
which we used to prove Property (3) (we denote $w_1$ as $w$).
\begin{align}
F_{1,n,1}(\{ u \}_n|w|1)
|_{w=t^{-1}u_n}
=&
(1-t)(u_n^2-1) \prod_{j=1}^{n-1}(tu_j-u_n)(u_j u_n-1) \nonumber \\
&\times \frac{1}{(n-1)!}
\sum_{\sigma^\prime \in S_{n-1}}
\prod_{2 \le j < k \le n}(u_{\sigma^\prime(j)}u_{\sigma^\prime(k)}-t)
\prod_{j=2}^n (Bu_{\sigma^\prime(j)}-A).
\label{righthandsideaftersubstitutionthree}
\end{align}
Since
\begin{align}
\prod_{2 \le j < k \le n}(u_{\sigma^\prime(j)}u_{\sigma^\prime(k)}-t)
=&\prod_{1 \le j < k \le n-1}(u_j u_k-t), \\
\prod_{j=2}^n (Bu_{\sigma^\prime(j)}-A)
=&\prod_{j=1}^{n-1}(Bu_j-A),
\end{align}
the summands are all equal,
and the sum in the right hand side of
\eqref{righthandsideaftersubstitutionthree} becomes 
$\displaystyle (n-1)!
\prod_{1 \le j < k \le n-1}(u_j u_k-t)
\prod_{j=1}^{n-1}(Bu_j-A)$.
Thus, we have the complete factorization
\begin{align}
&F_{1,n,1}(\{ u \}_n|w|1)|_{w=t^{-1}u_n} \nonumber \\
=&
(1-t)(u_n^2-1) \prod_{j=1}^{n-1}(tu_j-u_n)(u_j u_n-1) \nonumber \\
&\times \frac{1}{(n-1)!}
(n-1)!
\prod_{1 \le j < k \le n-1}(u_j u_k-t)
\prod_{j=1}^{n-1}(Bu_j-A) \nonumber \\
=&(1-t)\prod_{j=1}^{n-1}(tu_j-u_n)(Bu_j-A)
\prod_{j=1}^n(u_j u_n-1)
\prod_{1 \le j < k \le n-1}(u_j u_k-t),
\end{align}
which implies that $F_{1,n,1}(\{ u \}_n|w|1)$
satisfies the same property with $W_{1,n,1}(\{ u \}_n|w|1)$.
Hence, Property (4) for the case $m=1$ is proved.

The case $m=0$ can be shown in a similar way.
Writing down the case $N=1, \ m=0$
of \eqref{righthandsideaftersubstitutiontwo},
one immediately sees that all the summands are equal
and we have
\begin{align}
F_{1,n,0}(\{ u \}_n|w)
&=\frac{1}{n!} \sum_{\sigma \in S_n}
\prod_{j=1}^n (u_{\sigma(j)}-tw)
\prod_{1 \le j < k \le n} (u_{\sigma(j)} u_{\sigma(k)}-t)
\prod_{j=1}^n (Bu_{\sigma(j)}-A) \nonumber \\
&=\frac{1}{n!} n!
\prod_{j=1}^n (u_{j}-tw)
\prod_{1 \le j < k \le n} (u_{j} u_{k}-t)
\prod_{j=1}^n (Bu_{j}-A) \nonumber \\
&=\prod_{j=1}^n (u_{j}-tw)
\prod_{1 \le j < k \le n} (u_{j} u_{k}-t)
\prod_{j=1}^n (Bu_{j}-A),
\end{align}
which shows that $F_{1,n,0}(\{ u \}_n|w)$
is the same with $W_{1,n,0}(\{ u \}_n|w)$.

We have proved that the symmetric function
$F_{N,n,m}(u_1,\dots,u_n|w_1,\dots,w_N|x_1,\dots,x_m)$
satisfies all the
Properties (1), (2), (3) and (4) in
Proposition \ref{propertiesfordomainwallboundarypartitionfunction},
hence it is the explicit form of the
wavefunctions with triangular boundary
$Z_{N,n,m}(u_1,\dots,u_n|w_1,\dots,w_N|x_1,\dots,x_m)=
F_{N,n,m}(u_1,\dots,u_n|w_1,\dots,w_N|x_1,\dots,x_m)$.
\end{proof}

\section{Conclusion}
In this paper, we introduced an analogue of the wavefunctions
of an integrable six-vertex model with triangular boundary.
We used the $U_q(sl_2)$ $R$-matrix as the bulk weights
and a triangular $K$-matrix as the boundary weights.
We extended our recent work \cite{Motr} of the
Izergin-Korepin analysis on the domain wall boundary
partition function to the wavefunctions
and listed the properties about the degree, symmetry, two recursion relations
and two initial conditions.
We proved that a class of certain explicit symmetric functions
satisfies all the required properties, which means that
the symmetric functions is nothing but the explicit form
which represents the wavefunctions with triangular boundary.

The main motivation for introducing and studying the wavefunctions
with triangular boundary is that 
there are similar combinatorial objects
in algebraic combinatorics and Schubert calculus
such as the non-intersecting lattice paths and
excited Young diagrams \cite{St,Ok,IN1,IN2,HK}.
Some of those combinatorial objects can be regarded as
the wavefunctions of integrable lattice models with triangular boundary,
with the bulk weights given by an $R$-matrix of an
integrable five-vertex model,
and the boundary weights given by a triangular $K$-matrix.
Our result can be regarded as an extension of these combinatorial objects
by using the six-vertex $U_q(sl_2)$ $R$-matrix as the bulk weights,
and more general triangular $K$-matrix as the boundary weights.
The ordinary wavefunctions \cite{MS,Motegi} is represented by the
Grothendieck polynomials of type $A$ Grassmannian variety
and their quantum group deformations.
See the Appendix for a proof of the ordinarywavefunctions
based on the Izergin-Korepin analysis.

We called the class of partition functions treated in this paper
as ``the wavefunctions''.
For the case of ordinary wavefunctions, it is the
inner product between the state vector of spins
and the off-shell Bethe vector,
which becomes the eigenvector of the XXZ spin chain
when the Bethe ansatz equations
are imposed on the spectral parameters.
However, we do not know at this moment
if the ``off-shell'' Bethe vector
for the case of triangular boundary becomes
an eigenfunction of some Hamiltonian or transfer matrix.
This is the reason why we sometimes called the
partition functions treated in this paper as
``an analogue of the wavefunctions''.
It seems worthwhile to investigate whether such Hamiltonian
or transfer matrix exist, and if so, what are their explicit forms.

Extending the analysis to various integrable models
such as the higher rank models, elliptic models is also an
interesting topic.
Formulating the wavefunctions of elliptic integrable models \cite{eightvertex}
by using the dynamical $R$-matrix as the bulk weights
and the triangular elliptic $K$-matrix \cite{DG,IK} as the boundary weights,
using the Izergin-Korepin analysis
and finding the corresponding elliptic symmetric functions
with the help of complex analysis is an interesting work.
See \cite{Ros,PRS,FK,YZ,Chinesegroup,Chinesegroup2,Galleasone,Galleastwo,Galleasthree}
for examples on the treatment of the domain wall boundary partition functions
and scalar products of the elliptic and trigonometric face models
by using various methods developed to analzye those objects.
Also interesting topics are to study wavefunctions constructed
from the $R$- and $K$-matrices of higher dimensional representations.
See \cite{Ga} for example on this direction.

\section*{Acknowledgements}
This work was partially supported by grant-in-Aid
for Research Activity start-up No. 15H06218
and Scientific Research (C) No. 16K05468.

\appendix
\def\thesection{\Alph{section}}
\def\reference{\relax\refpar}

\section{Ordinary wavefunctions}

\begin{figure}[ht]
\includegraphics[width=12cm]{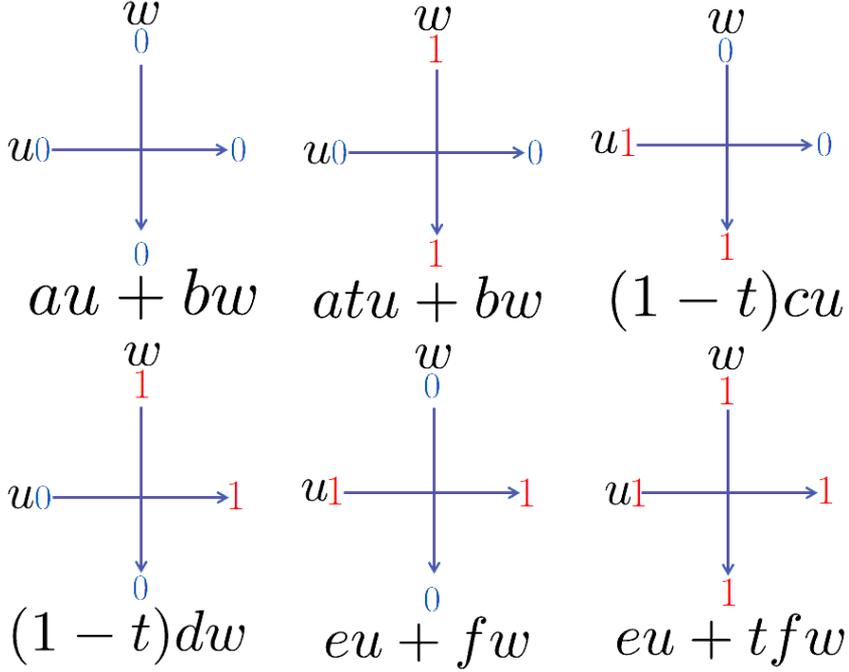}
\caption{The $L$-operator $L(u,w)$ \eqref{generalizedloperator}.
The horizontal line is the space $W$, and the vertical line
is the space $V$.
}
\label{picturegeneralizedloperator}
\end{figure}

We give here the Izergin-Korepin analysis
on the ordinary wavefunctions in this Appendix.
We remark that the result presented here
can be obtained by applying the general result on the 
ordinary wavefunctions, for example \cite{Borodin,BP1},
to the following $L$-operator presented below.
We present a proof based on the Izergin-Korepin analysis here
since it is an illustrative example.

We analyze the wavefunctions constructed from the 
following generalized $L$-operator \cite{MS2}
(Figure \ref{picturegeneralizedloperator})
\begin{eqnarray}
L_{aj}(u,w,a,b,c,d,e,f)=\left( 
\begin{array}{cccc}
au+bw & 0 & 0 & 0 \\
0 & atu+bw & (1-t)cu & 0 \\
0 & (1-t)dw & eu+fw & 0 \\
0 & 0 & 0 & eu+tfw
\end{array}
\right), \label{generalizedloperator}
\end{eqnarray}
acting on the tensor product $W_a \otimes V_j$
where $V_j$ is also a complex two-dimensional vector space,
and the parameters
$a$, $b$, $c$, $d$, $e$ and $f$ are constant parameters
(do not depend on the spectral parameter $u$)
and must obey the following relations
\begin{align}
(1-t)cd+af-be=0, \ (t^2-t)cd+t^2 af-be=0. \label{constraints}
\end{align}
If one assumes $t \neq 1$, the relations \eqref{constraints}
further reduce to
\begin{align}
cd+af=0, \ tcd+be=0. \label{constraints2}
\end{align}
We frequently abbreviate $L_{aj}(u,w,a,b,c,d,e,f)$
as $L_{aj}(u,w)$ for simplicity.
The matrix elements of the $L$-operator \eqref{generalizedloperator}
is explicitly given by
\begin{align}
{}_a \langle 0| {}_j \langle 0 | L_{a j}(u,w)
|0 \rangle_a | 0 \rangle_j&=au+bw, \\
{}_a \langle 0| {}_j \langle 1 | L_{a j}(u,w)
|0 \rangle_a | 1 \rangle_j&=atu+bw, \\
{}_a \langle 0| {}_j \langle 1 | L_{a j}(u,w)
|1 \rangle_a | 0 \rangle_j&=(1-t)cu, \\
{}_a \langle 1| {}_j \langle 0 | L_{a j}(u,w)
|0 \rangle_a |1 \rangle_j&=(1-t)dw, \\
{}_a \langle 1| {}_j \langle 0 | L_{a j}(u,w)
|1 \rangle_a | 0 \rangle_j&=eu+fw, \\
{}_a \langle 1| {}_j \langle 1 | L_{a j}(u,w)
|1 \rangle_a | 1 \rangle_j&=eu+ftw.
\end{align}

\begin{figure}[ht]
\includegraphics[width=12cm]{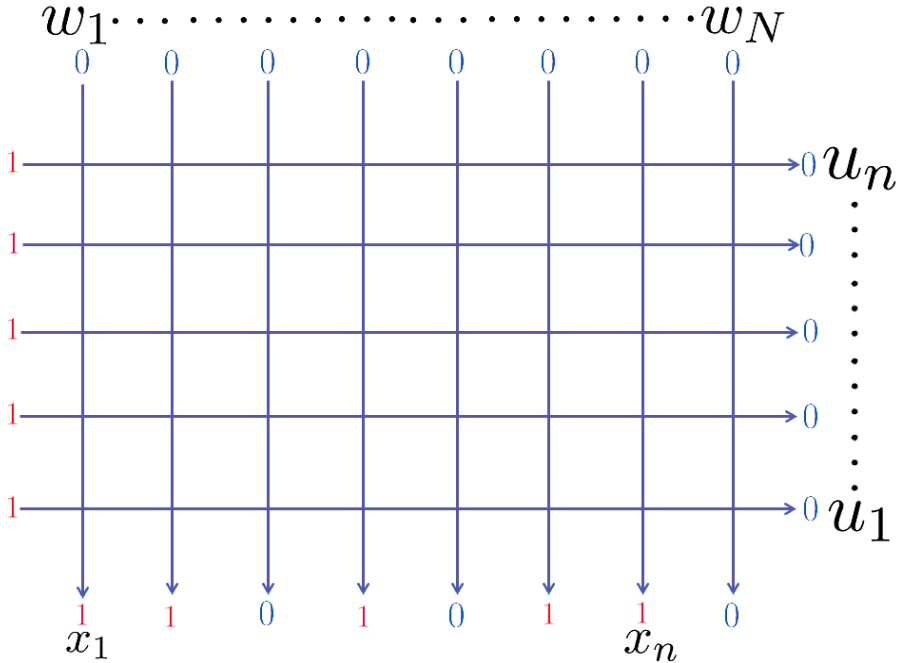}
\caption{The ordinary wavefunctions
$OW_{N,n}(u_1,\dots,u_n|w_1,\dots,w_N|x_1,\dots,x_n)$
\eqref{ordinaryDWBPF}.
This figure illustrates the case $N=8, \ n=5, \ x_1=1, \ x_2=2, \ x_3=4,
\ x_4=6, \ x_5=7$.
}
\label{ordinarypictureDWBPF}
\end{figure}

\begin{figure}[ht]
\includegraphics[width=12cm]{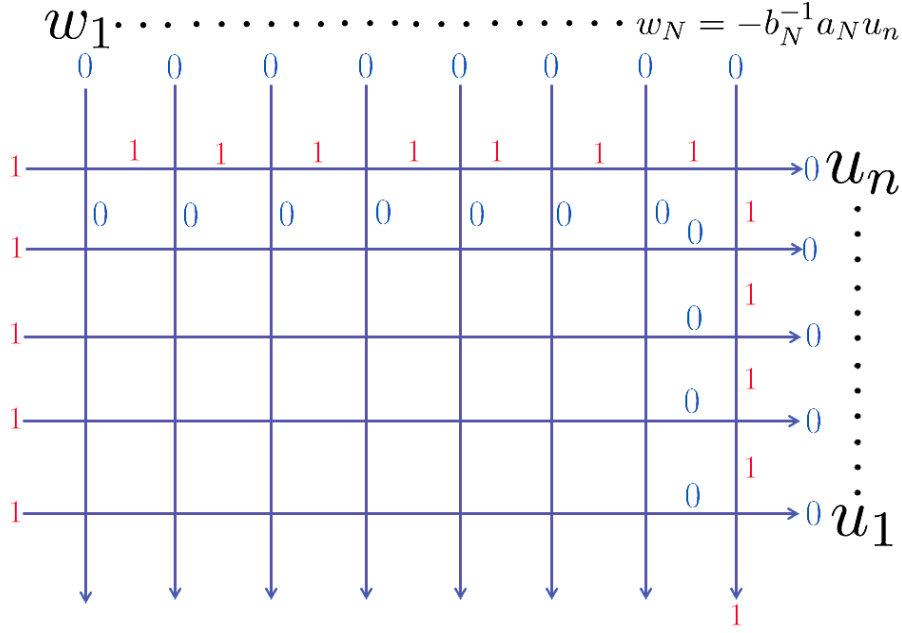}
\caption{The recursion relation
$OW_{N,n}(u_1,\dots,u_n|w_1,\dots,w_N|x_1,\dots,x_n)$,
$x_n=N$ evaluated at $w_N=-b_N^{-1}a_N u_n$
\eqref{ordinaryrecursionwavefunction}
.}
\label{ordinarypicturerecursion}
\end{figure}

\begin{figure}[ht]
\includegraphics[width=12cm]{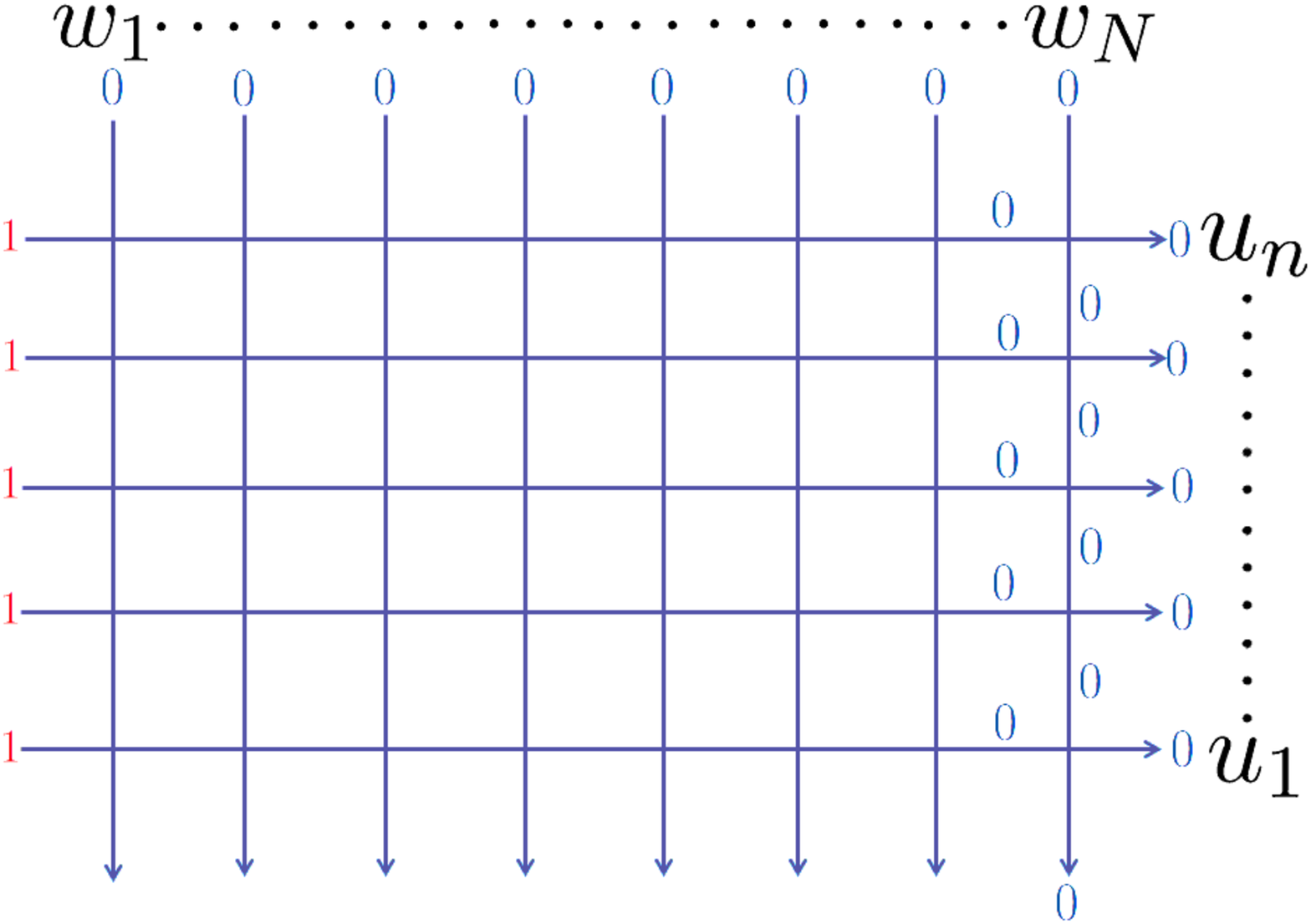}
\caption{The factorization of
$OW_{N,n}(u_1,\dots,u_n|w_1,\dots,w_N|x_1,\dots,x_n)$,
$x_n \neq N$
\eqref{ordinaryrecursionwavefunction2}
.}
\label{ordinarypicturerecursion2}
\end{figure}

The $L$-operator \eqref{generalizedloperator}
together with the $U_q(sl_2)$ $R$-matrix \eqref{rmatrix}
satisfies the $RLL$-type Yang-Baxter relation
\begin{align}
R_{ab}(u_1,u_2)L_{aj}(u_1,w)L_{bj}(u_2,w)
=L_{bj}(u_2,w)L_{aj}(u_1,w)R_{ab}(u_1,u_2), \label{RLL}
\end{align}
acting on $W_a \otimes W_b \otimes V_j$.
\eqref{generalizedloperator} and \eqref{RLL}
contains the $U_q(sl_2)$ $R$-matrix \eqref{rmatrix}
and the $RRR$-type Yang-Baxter relation \eqref{yangbaxter}
as a special case
$a=1$, $b=-t$, $c=1$, $d=1$, $e=1$, $f=-1$.

In the quantum inverse scattering method,
what we first do is to
construct the monodromy matrix $T_a(u|w_1,\dots,w_N)$ from the $L$-operator as
\begin{align}
T_{a}(u|w_1,\dots,w_N)&=L_{a N}(u,w_N) \cdots L_{a 1}(u,w_1)
\nonumber \\
&=
\begin{pmatrix}
A(u|w_1,\dots,w_N) & B(u|w_1,\dots,w_N)  \\
C(u|w_1,\dots,w_N) & D(u|w_1,\dots,w_N)
\end{pmatrix}_{a} \in \mathrm{End}(W_a \otimes V_1 \otimes \cdots \otimes V_N).
\label{monodromy}
\end{align}
Here, 
$L_{a j}(u,w_j)=L_{a_N}(u,w_j,a_j,b_j,c_j,d_j,e_j,f_j)$, $j=1,\dots,N$
is the $L$-operator where $a_j,b_j,c_j,d_j,e_j$ and $f_j$ satisfies
\begin{align}
c_j d_j+a_j f_j=0, \ tc_j d_j+b_j e_j=0,
\end{align}
for each $j$.

The matrix elements of the monodromy matrix
\begin{align}
A(u|w_1,\dots,w_N)=_a \langle 0 |T_a(u|w_1,\dots,w_N)| 0 \rangle_a, \\
B(u|w_1,\dots,w_N)=_a \langle 0 |T_a(u|w_1,\dots,w_N)| 1 \rangle_a, \\
C(u|w_1,\dots,w_N)=_a \langle 1 |T_a(u|w_1,\dots,w_N)| 0 \rangle_a, \\
D(u|w_1,\dots,w_N)=_a \langle 1 |T_a(u|w_1,\dots,w_N)| 1 \rangle_a,
\end{align}
are $2^N \times 2^N$ matrices
acting on the tensor product of the quantum spaces
$V_1\otimes \dots \otimes V_N$.

Particularly important is the $B$-operator
which has the role of creating down spins in the quantum spaces
$V_1 \otimes \cdots \otimes V_N$.
We next introduce the following state vector \\
$|\Phi_{N,n}(u_1,\dots,u_n|w_1,\dots,w_N) \rangle \in V_1 \otimes \cdots \otimes V_N$ using the $B$-operators as
\begin{align}
|\Phi_{N,n}(u_1,\dots,u_n|w_1,\dots,w_N)
\rangle=B(u_1|w_1,\dots,w_N) \cdots B(u_n|w_1,\dots,w_N)|\Omega \rangle_N,
\label{ordinarystatevector}
\end{align}
where $|\Omega \rangle_N:=|0 \rangle_1 \otimes \cdots \otimes |0 \rangle_N
\in V_1 \otimes \cdots \otimes V_N$
is the vacuum state in the tensor product of quantum spaces.
The state vector \eqref{ordinarystatevector}
is sometimes called as the off-shell Bethe vector.
If one imposes certain constraints on the spectral parameters $\{ u_j \}$
called the Bethe ansatz equations,
the state vector \eqref{ordinarystatevector} becomes the eigenvectors
(Bethe vectors) of the transfer matrix $A(u|w_1,\dots,w_N)+D(u|w_1,\dots,w_N)$
which is a generating function of operators including the Hamiltonian.

Due to the ice rule, each $B$-operator creates one down spin
in the quantum spaces. This fact and that the state vector
\eqref{ordinarystatevector} is constructed from $N$-layers
of the $B$-operators acting on the vacuum state
$|\Omega \rangle_N$, the state vector \eqref{ordinarystatevector}
is an $N$-down spin state.
To construct a nonvanishing inner product,
we introduce the dual $n$-down spin state
\begin{align}
\langle x_1 \cdots x_n|
&=(_1 \langle 0| \otimes \cdots \otimes {}_N \langle 0|)
\prod_{j=1}^n \sigma^+_{x_j}
\in V_1^* \otimes \cdots \otimes V_N^*
, \label{ordinarydualparticleconfiguration}
\end{align}
which are states labelling the configurations
of down spins
$1 \le x_1 < x_2 < \cdots < x_n \le N$.

The ordinary wavefunctions
$OW_{N,n}(u_1,\dots,u_n|w_1,\dots,w_N|x_1,\dots,x_n)$
is defined as the inner product between the state vector
$|\Phi_{N,n}(u_1,\dots,u_n|w_1,\dots,w_N) \rangle$
and the $n$-down spin state $\langle x_1 \cdots x_n|$
\begin{align}
OW_{N,n}(u_1,\dots,u_n|w_1,\dots,w_N|x_1,\dots,x_n)
=\langle x_1 \cdots x_n|\Phi_{N,n}(u_1,\dots,u_n|w_1,\dots,w_N) \rangle.
\label{ordinaryDWBPF}
\end{align}
See Figure \ref{ordinarypictureDWBPF} for a pictorial description
of \eqref{ordinaryDWBPF}.

We list the properties
of the ordinary wavefunctions \\
$OW_{N,n}(u_1,\dots,u_n|w_1,\dots,w_N|x_1,\dots,x_n)$
by the Izergin-Korepin analysis.

\begin{proposition} 
\label{ordinarypropertiesfordomainwallboundarypartitionfunction}
The wavefunctions
$OW_{N,n}(u_1,\dots,u_n|w_1,\dots,w_N|x_1,\dots,x_n)$
satisfies the following properties. \\
\\
 (1) $OW_{N,n}(u_1,\dots,u_n|w_1,\dots,w_N|x_1,\dots,x_n)$
is a polynomial of degree $n-1$ in $w_N$
if $x_n=N$ and degree $n$ if $x_n \neq N$.
\\
 (2) $OW_{N,n}(u_1,\dots,u_n|w_1,\dots,w_N|x_1,\dots,x_n)$ is symmetric
with respect to $u_j$, $j=1,\dots,n$.
\\
(3) The following recursive relations between the
wavefunctions hold if $x_n=N$
(Figure \ref{ordinarypicturerecursion}):
\begin{align}
&OW_{N,n}(u_1,\dots,u_n|w_1,\dots,w_N|x_1,\dots,x_n)
|_{w_N=-b_N^{-1}a_N^{-1}u_n}
\nonumber \\
=&(1-t)c_N u_n \prod_{j=1}^{n-1} a_N(tu_j-u_n)
\prod_{j=1}^{N-1}(e_j u_n+f_j w_j)
\nonumber \\
&\times OW_{N-1,n-1}(u_1,\dots,u_{n-1}|w_1,\dots,w_{N-1}|x_1,\dots,x_{n-1})
. \label{ordinaryrecursionwavefunction}
\end{align}
If $x_n \neq N$, the following factorizations hold for the wavefunctions
(Figure \ref{ordinarypicturerecursion2}):
\begin{align}
&OW_{N,n}(u_1,\dots,u_n|w_1,\dots,w_N|x_1,\dots,x_n) \nonumber \\
=&\prod_{j=1}^n (a_N u_j+b_N w_N)
OW_{N-1,n}(u_1,\dots,u_n|w_1,\dots,w_{N-1}|x_1,\dots,x_n).
\label{ordinaryrecursionwavefunction2}
\end{align}
\\
(4) The following holds for the case $N=n=1$
\begin{align}
&
OW_{1,1}(u|w|1)=(1-t)cu.
\label{ordinaryinitialrecursion}
\end{align}
\end{proposition}

\begin{proof}
Properties (1), (2), (3) and (4)
can be shown in a similar way
with the wavefunctions with triangular boundary
or the ordinary domain wall boundary partition functions.
We give brief explanations below for each Properties.

Property (3) can be shown by using the graphical
representation of the ordinary wavefunctions,
${}_{a} \langle 0| {}_{N} \langle 0|
L_{aN}(u_n,-b_N^{-1}a_N^{-1}u_n,a_N,b_N,c_N,d_N,e_N,f_N)|0 \rangle_a |0 \rangle_N=0$
and the ice rule.
See Figures \ref{ordinarypicturerecursion}
and \ref{ordinarypicturerecursion2}
which explains \eqref{ordinaryrecursionwavefunction}
and \eqref{ordinaryrecursionwavefunction2} respectively.
The factorization of the wavefunctions
\eqref{ordinaryrecursionwavefunction2} also shows
Property (1) for the case $x_n \neq N$.

Property (1) for the case $x_n=N$ can be shown
by inserting the completeness relation
in one spin down state sector.

Property (2) can be shown by the railroad argument
using the Yang-Baxter relation.
Equivalently, it follows from the commutativity
of the $B$-matrix
\begin{align}
{[}B(u_1|w_1,\dots,w_N), B(u_2|w_1,\dots,w_N) {]}=0.
\end{align}
This commutativity follows from writing down
a matrix element of the equality of the
intertwining relation between the monodromy matrices
\begin{align}
R_{ab}(u_1,u_2)T_{a}(u_1,w)T_{b}(u_2,w)
=T_{b}(u_2,w)T_{a}(u_1,w)R_{ab}(u_1,u_2),
\end{align}
which can be obtained by using the $RLL$ relation \eqref{RLL} repeatedly.

Property (4) is obvious since
$OW_{1,1}(u|w|1)={}_{a} \langle 0| {}_{1} \langle 1|
L_{a1}(u,w,a,b,c,d,e,f)|1 \rangle_a |0 \rangle_1=(1-t)cu$.
\end{proof}

\begin{definition}
We define the following symmetric function \\
$OF_{N,n}(u_1,\dots,u_n|w_1,\dots,w_N|x_1,\dots,x_n)$
which depends on the symmetric variables $u_1,\dots,u_n$,
complex parameters $w_1,\dots,w_N$
and integers $x_1,\dots,x_n$ satisfying
$1 \le x_1 < \cdots < x_n \le N$,
\begin{align}
&OF_{N,n}(u_1,\dots,u_n|w_1,\dots,w_N|x_1,\dots,x_n) \nonumber \\
=
&
\sum_{\sigma \in S_n}
\prod_{j=1}^n \prod_{k=x_j+1}^N (a_k u_{\sigma(j)}+b_k w_k)
\prod_{1 \le j < k \le n}
\frac{tu_{\sigma(j)}-u_{\sigma(k)}}{u_{\sigma(j)}-u_{\sigma(k)}}
\nonumber \\
&\times
\prod_{j=1}^n \prod_{k=1}^{x_j-1}(e_k u_{\sigma(j)}+f_k w_k)
\prod_{j=1}^n (1-t)c_{x_j}u_{\sigma(j)}.
\label{ordinaryrighthandside}
\end{align}
\end{definition}

We make a comment on the symmetric function
$OF_{N,n}(u_1,\dots,u_n|w_1,\dots,w_N|x_1,\dots,x_n)$.
If one takes the homogeneous limit $w_j=1$, $j=1,\dots,N$
and the following specialization
$a_j=1$, $b_j=t \beta$, $c_j=1$, $d_j=1$, $e_j=-\beta^{-1}$, $f_j=-1$,
$j=1,\dots,N$, the symmetric function
\eqref{ordinaryrighthandside} becomes
\begin{align}
&OF_{N,n}(u_1,\dots,u_n|1,\dots,1|x_1,\dots,x_n) \nonumber \\
=&\prod_{j=1}^n \frac{(1-t)u_j(u_j+t \beta)^N}{-\beta^{-1} u_j-1}
\prod_{1 \le j < k \le n} \frac{tu_j-u_k}{u_j-u_k} \nonumber \\
&\times \sum_{\sigma \in S_n}
\prod_{\substack{1 \le j<k \le n \\ \sigma(j)>\sigma(k)}}
\frac{u_{\sigma(k)}-tu_{\sigma(j)}}{tu_{\sigma(k)}-u_{\sigma(j)}}
\prod_{j=1}^n \Bigg(\frac{-\beta^{-1} u_{\sigma(j)}-1}{u_{\sigma(j)}+t \beta} \Bigg)^{x_j}. \label{homogeneouslimit}
\end{align}
If one furthermore set the parameter of the quantum group
$t$ to $t=0$, \eqref{homogeneouslimit}
essentially becomes the Grothendieck polynomials
\begin{align}
OF_{N,n}(u_1,\dots,u_n|1,\dots,1|x_1,\dots,x_n)|_{t=0}
=&(-\beta)^{-n(n-1)/2}\prod_{j=1}^n u_j^N 
G_\lambda(\bs{z};\beta). \label{correspondence}
\end{align}
Here, $G_{\lambda}(\bs{z};\beta)$ is the
$\beta$-Grothendieck polynomials of type $A$ Grassmannian variety
\cite{IN2,LS,FoKi,Buch,IS,Mc}, which is known to have the following
determinant form
\begin{align}
G_\lambda(\bs{z};\beta)=
   \frac{\mathrm{det}_n(z_j^{\lambda_k+n-k}(1+\beta z_j)^{k-1})}
        {\prod_{1 \le j < k \le n}(z_j-z_k)}.
 \label{GR}
\end{align}
In the correspondence \eqref{correspondence},
the symmetric variables $\bs{z}=\{z_1,\dots,z_n\}$
for the Grothendieck polynomials $G_\lambda(\bs{z};\beta)$ and the
spectral parameters $u_1,\dots,u_n$ of the symmetric functions
$OF_{N,n}(u_1,\dots,u_n|1,\dots,1|x_1,\dots,x_n)|_{t=0}$
are related by the correspondence
$z_j=-\beta^{-1}-u_j^{-1}$, $j=1,\dots,n$.
Also, the Young diagrams
$\lambda=(\lambda_1,\lambda_2,\dots,\lambda_n) \in \mathbb{Z}^n$
($N-n \ge \lambda_1 \ge \lambda_2 \ge \dots \ge \lambda_n \ge 0$)
in $G_\lambda(\bs{z};\beta)$
and the sequence of integers $x_1,\dots,x_n$ satisfying
$1 \le x_1 < \cdots < x_n \le N$
in $OF_{N,n}(u_1,\dots,u_n|1,\dots,1|x_1,\dots,x_n)|_{t=0}$
are connected by the translation rule
$\lambda_j=x_{n-j+1}-n+j-1$, $j=1,\dots,n$.

We have the following correspondence between the
ordinary wavefunctions
of the six-vertex model
and the symmetric function
$OF_{N,n}(u_1,\dots,u_n|w_1,\dots,w_N|x_1,\dots,x_n)$.

\begin{theorem}
The ordinary wavefunctions of the six-vertex model
\\
$OZ_{N,n}(u_1,\dots,u_n|w_1,\dots,w_N|x_1,\dots,x_n)$
is explicitly expressed as the
symmetric function
$OF_{N,n}(u_1,\dots,u_n|w_1,\dots,w_N|x_1,\dots,x_n)$
\begin{align}
OZ_{N,n}(u_1,\dots,u_n|w_1,\dots,w_N|x_1,\dots,x_n)=
OF_{N,n}(u_1,\dots,u_n|w_1,\dots,w_N|x_1,\dots,x_n).
\end{align}
\end{theorem}

\begin{proof}
We prove this theorem by showing that
the symmetric funcion
\eqref{ordinaryrighthandside} \\
$OF_{N,n}(u_1,\dots,u_n|w_1,\dots,w_N|x_1,\dots,x_n)$
satisfies all the four Properties in
Proposition \ref{ordinarypropertiesfordomainwallboundarypartitionfunction}.
The proof goes along the same line as the case with triangular boundary,
but is much simpler.

To show Property (1),
first note that the factor
$\displaystyle \prod_{j=1}^n \prod_{k=x_j+1}^N (a_k u_{\sigma(j)}+b_k w_k)$
in \\
$OF_{N,n}(u_1,\dots,u_n|w_1,\dots,w_N|x_1,\dots,x_n)$
is a polynomial of degree $n-1$ in $w_N$ if $x_n=N$
and degree $n$ if $x_n \neq N$.
We can also immediately
see that the dependence on $w_N$ just only comes from this factor,
hence Property (1) is proved.

It is also easy to see that
$OF_{N,n}(u_1,\dots,u_n|w_1,\dots,w_N|x_1,\dots,x_n)$
is symmetric with respect to $u_j$, $j=1,\dots,n$
from the fact that the sum is over all permutations
of the variables $u_j$, $j=1,\dots,n$.

Next we show Property (3).
We first prove the function
$OF_{N,n}(\{ u \}_n|w_1,\dots,w_N|x_1,\dots,x_n)$ satisfies
\eqref{ordinaryrecursionwavefunction}
for the case $x_n=N$.
In this case, we first note that the factor
\begin{align}
\prod_{j=1}^n \prod_{k=x_j+1}^N (a_k u_{\sigma(j)}+b_k w_k),
\end{align}
in each summand essentially becomes
\begin{align}
\prod_{j=1}^{n-1} \prod_{k=x_j+1}^N (a_k u_{\sigma(j)}+b_k w_k).
\label{ordinaryfactorconsideration}
\end{align}
Concentrating on the factor
$\displaystyle
\prod_{j=1}^{n-1} (a_N u_{\sigma(j)}+b_N w_N)
$ from \eqref{ordinaryfactorconsideration},
one finds this factor vanishes unless $\sigma$ satisfies $\sigma(n)=n$
if one substitutes $w_N=-b_N^{-1} a_N u_n$.

Therefore, only the summands satisfying $\sigma(n)=n$ 
in \eqref{ordinaryrighthandside} survive
after the substitution $w_N=-b_N^{-1} a_N u_n$.
Keeping this in mind, one rewrites
$OF_{N,n}(\{ u \}_n|w_1,\dots,w_N|x_1,\dots,x_n)$
evaluated at $w_N=-b_N^{-1} a_N u_n$
by using the symmetric group $S_{n-1}$
where every $\sigma^\prime \in S_{n-1}$ satisfies
$\{\sigma^\prime(1),\cdots,\sigma^\prime(n-1)\}=\{1,\cdots,n-1 \}$ as follows:
\begin{align}
&OF_{N,n}(u_1,\dots,u_n|w_1,\dots,w_N|x_1,\dots,x_n)
|_{w_N=-b_N^{-1} a_N u_n} \nonumber \\
=&
\sum_{\sigma^\prime \in S_{n-1}}
\prod_{j=1}^{n-1} \prod_{k=x_j+1}^{N-1}(a_k u_{\sigma^\prime(j)}+b_k w_k)
\prod_{j=1}^{n-1}a_N(u_{\sigma^\prime(j)}-u_n)
\nonumber \\
&\times \prod_{1 \le j < k \le n-1}
\frac{tu_{\sigma^\prime(j)}-u_{\sigma^\prime(k)}}{u_{\sigma^\prime(j)}
-u_{\sigma^\prime(k)}}
\prod_{j=1}^{n-1}
\frac{tu_{\sigma^\prime(j)}-u_n}{u_{\sigma^\prime(j)}
-u_n}
\prod_{j=1}^{n-1} \prod_{k=1}^{x_j-1}(e_k u_{\sigma^\prime(j)}+f_k w_k)
\prod_{k=1}^{N-1}(e_k u_n+f_k w_k) \nonumber \\
&\times
(1-t)c_N u_n
\prod_{j=1}^{n-1} (1-t)c_{x_j} u_{\sigma^\prime(j)}.
\label{ordinaryrighthandsideaftersubstitution}
\end{align}
One easily notes that
the factors $\displaystyle
\prod_{k=1}^{N-1}(e_k u_n+f_k w_k)$
and $(1-t)c_N u_n$ in the sum are independent of the permutation
$S^\prime_{n-1}$.
One also finds the factor
$\displaystyle \prod_{j=1}^{n-1}\frac{t u_{\sigma^\prime(j)}-u_n}
{u_{\sigma^\prime(j)}-u_n}
\prod_{j=1}^{n-1} a_N (u_{\sigma^\prime(j)}-u_n)$
can be simplified as
\begin{align}
&\prod_{j=1}^{n-1}\frac{t u_{\sigma^\prime(j)}-u_n}
{u_{\sigma^\prime(j)}-u_n}
\prod_{j=1}^{n-1} a_N (u_{\sigma^\prime(j)}-u_n)
=\prod_{j=1}^{n-1}a_N (tu_{\sigma^\prime}(j)-u_n)
=\prod_{j=1}^{n-1}a_N (tu_j-u_n).
\end{align}
Thus, \eqref{ordinaryrighthandsideaftersubstitution}
can be rewritten furthermore as
\begin{align}
&OF_{N,n}(u_1,\dots,u_n|w_1,\dots,w_N|x_1,\dots,x_n)|_{w_N=-b_N^{-1} a_N u_n} \nonumber \\
=&
(1-t)c_N u_n \prod_{j=1}^{n-1}a_N (tu_j-u_n) \prod_{j=1}^{N-1}(e_j u_n+f_j w_j)
\sum_{\sigma^\prime \in S_{n-1}}
\prod_{j=1}^{n-1} \prod_{k=x_j+1}^{N-1}(a_k u_{\sigma^\prime(j)}+b_k w_k)
\nonumber \\
&\times \prod_{1 \le j < k \le n-1}
\frac{tu_{\sigma^\prime(j)}-u_{\sigma^\prime(k)}}{u_{\sigma^\prime(j)}
-u_{\sigma^\prime(k)}}
\prod_{j=1}^{n-1} \prod_{k=1}^{x_j-1}(e_k u_{\sigma^\prime(j)}+f_k w_k)
\prod_{j=1}^{n-1} (1-t)c_{x_j} u_{\sigma^\prime(j)}.
\label{ordinaryrighthandsideaftersubstitutiontwo}
\end{align}
Since
\begin{align}
&\sum_{\sigma^\prime \in S_{n-1}}
\prod_{j=1}^{n-1} \prod_{k=x_j+1}^{N-1}(a_k u_{\sigma^\prime(j)}+b_k w_k)
\nonumber \\
&\times \prod_{1 \le j < k \le n-1}
\frac{tu_{\sigma^\prime(j)}-u_{\sigma^\prime(k)}}{u_{\sigma^\prime(j)}
-u_{\sigma^\prime(k)}}
\prod_{j=1}^{n-1} \prod_{k=1}^{x_j-1}(e_k u_{\sigma^\prime(j)}+f_k w_k)
\prod_{j=1}^{n-1} (1-t)c_{x_j} u_{\sigma^\prime(j)}
\nonumber \\
=&
OF_{N-1,n-1}(u_1,\dots,u_{n-1}|w_1,\dots,w_{N-1}|x_1,\dots,x_{n-1})
,
\end{align}
one finds that
\eqref{ordinaryrighthandsideaftersubstitutiontwo} is nothing but the
following recursion relation
for the symmetric function
$OF_{N,n}(u_1,\dots,u_n|w_1,\dots,w_N|x_1,\dots,x_n)$
\begin{align}
&OF_{N,n}(u_1,\dots,u_n|w_1,\dots,w_N|x_1,\dots,x_n)
|_{w_N=-b_N^{-1}a_N^{-1}u_n}
\nonumber \\
=&(1-t)c_N u_n \prod_{j=1}^{n-1} a_N(tu_j-u_n)
\prod_{j=1}^{N-1}(e_j u_n+f_j w_j)
\nonumber \\
&\times OF_{N-1,n-1}(u_1,\dots,u_{n-1}|w_1,\dots,w_{N-1}|x_1,\dots,x_{n-1}),
\end{align}
which is exactly the same recursion relation the wavefunctions
\\
$OW_{N,n}(u_1,\dots,u_n|w_1,\dots,w_N|x_1,\dots,x_n)$ must satisfy.
Hence, Property (3) for the case $x_n=N$ is proved.

The case $x_n \neq N$ can be shown in a similar but much simpler way.
Rewriting \\
$OF_{N,n}(u_1,\dots,u_n|w_1,\dots,w_N|x_1,\dots,x_n)$ as
\begin{align}
&OF_{N,n}(u_1,\dots,u_n|w_1,\dots,w_N|x_1,\dots,x_n) \nonumber \\
=
&
\sum_{\sigma \in S_n}
\prod_{j=1}^n \prod_{k=x_j+1}^{N-1} (a_k u_{\sigma(j)}+b_k w_k)
\prod_{j=1}^n (a_N u_{\sigma(j)}+b_N w_N) \nonumber \\
&\times \prod_{1 \le j < k \le n}
\frac{tu_{\sigma(j)}-u_{\sigma(k)}}{u_{\sigma(j)}
-u_{\sigma(k)}}
\prod_{j=1}^n \prod_{k=1}^{x_j-1} (e_k u_{\sigma(j)}+f_k w_k)
\prod_{j=1}^n (1-t)c_{x_j}u_{\sigma(j)},
\label{originalrighthandsidenew}
\end{align}
and noting
\begin{align}
\displaystyle \prod_{j=1}^n (a_N u_{\sigma(j)}+b_N w_N)
=\prod_{j=1}^n (a_N u_j+b_N w_N),
\end{align}
we can take this factor out of the sum in
\eqref{originalrighthandsidenew} and we have
\begin{align}
&OF_{N,n}(u_1,\dots,u_n|w_1,\dots,w_N|x_1,\dots,x_n) \nonumber \\
=
&\prod_{j=1}^n (a_N u_j+b_N w_N)
\sum_{\sigma \in S_n}
\prod_{j=1}^n \prod_{k=x_j+1}^{N-1} (a_k u_{\sigma(j)}+b_k w_k)
\nonumber \\
&\times \prod_{1 \le j < k \le n}
\frac{tu_{\sigma(j)}-u_{\sigma(k)}}{u_{\sigma(j)}
-u_{\sigma(k)}}
\prod_{j=1}^n \prod_{k=1}^{x_j-1} (e_k u_{\sigma(j)}+f_k w_k)
\prod_{j=1}^n (1-t)c_{x_j}u_{\sigma(j)} \nonumber \\
=&\prod_{j=1}^n (a_N u_j+b_N w_N)
OF_{N-1,n}(u_1,\dots,u_n|w_1,\dots,w_{N-1}|x_1,\dots,x_n),
\end{align}
which is also exactly the recursion relation the ordinary wavefunctions \\
$OF_{N,n}(u_1,\dots,u_n|w_1,\dots,w_N|x_1,\dots,x_n)$
must satisfy for the case $x_n \neq N$.

It is trivial to check that
$OW_{N,n}(u_1,\dots,u_n|w_1,\dots,w_N|x_1,\dots,x_n)$ satisfies
$OW_{1,1}(u|w|1)=(1-t)cu$, hence Property (4) is shown.

Since we have proved that the symmetric function
$OF_{N,n}(u_1,\dots,u_n|w_1,\dots,w_N|x_1,\dots,x_n)$
satisfies all the
Properties (1), (2), (3) and (4) in
Proposition \ref{ordinarypropertiesfordomainwallboundarypartitionfunction},
we conclude it is the explicit form of the
ordinary wavefunctions \\
$OZ_{N,n}(u_1,\dots,u_n|w_1,\dots,w_N|x_1,\dots,x_n)=
OF_{N,n}(u_1,\dots,u_n|w_1,\dots,w_N|x_1,\dots,x_n)$.
\end{proof}

\end{document}